\documentclass[twocolumn,aps,showpacs,amssymb,floatfix]{revtex4}

\usepackage{graphicx}
\usepackage{bm}
\usepackage{epsf}
\usepackage{epsfig}
\usepackage{amsmath}
\usepackage{amssymb}
\usepackage{amsfonts}
\usepackage{float}
\usepackage{hyperref}
\usepackage{dsfont}  
\usepackage{slashed}  
\usepackage{booktabs}
\usepackage[sort&compress]{natbib}

\newcommand{\be}{\begin{equation}}  
\newcommand{\ee}{\end{equation}}  
\newcommand{\beq}{\begin{eqnarray}}  
\newcommand{\eeq}{\end{eqnarray}}
\newcommand{\Dsla}{\slashed{D}}
\renewcommand{\thefigure}{\arabic{figure}}

\begin{document}

\title{Evaluation of disconnected quark loops for hadron structure using GPUs}

\date{\today}
\author{C.~Alexandrou~$^{(a,b)}$\footnote{email:alexand@ucy.ac.cy}, M. Constantinou~$^{(a)}$, V. Drach~$^{(c)}$, K. Hadjiyiannakou~$^{(a)}$, K.~Jansen~$^{(a,c)}$, G. Koutsou~$^{(b)}$, A. Strelchenko~$^{(b)}$\footnote{Present address: Scientific Computing Division, Fermilab, Batavia, IL 60510-5011, USA}, A. Vaquero~$^{(b)}$}
\affiliation{$^{(a)}$ Department of Physics, University of Cyprus, P.O. Box 20537, 1678 Nicosia, Cyprus\\  
 $^{(b)}$ Computation-based Science and Technology Research  
    Center, The Cyprus Institute, 20 Kavafi Str., Nicosia 2121, Cyprus \\  
$^{(c)}$ NIC, DESY, Platanenallee 6, D-15738 Zeuthen, Germany}

\begin{abstract}
  
A number of stochastic methods developed for the calculation of
fermion loops are investigated and compared, in particular with
respect to their efficiency when implemented on Graphics Processing
Units (GPUs). We assess the performance of the various methods by
studying the convergence and statistical accuracy obtained for
observables that require a large number of stochastic noise vectors,
such as the isoscalar nucleon axial charge. The various methods are
also examined for the evaluation of sigma-terms where noise reduction
techniques specific to the twisted mass formulation can be utilized
thus reducing the required number of stochastic noise vectors.
\end{abstract}  


\maketitle 

\setcounter{figure}{\arabic{figure}}

\newcommand{\twopt}[5]{\langle G_{#1}^{#2}(#3;\mathbf{#4};\Gamma_{#5})\rangle}
\newcommand{\threept}[7]{\langle G_{#1}^{#2}(#3,#4;\mathbf{#5},\mathbf{#6};\Gamma_{#7})\rangle}  
  
\newcommand{\Op}{\mathcal{O}} 
\newcommand{\C}{\mathcal{C}} 
\newcommand{\eins}{\mathds{1}} 

\bibliographystyle{apsrev}

\section{Introduction}

The evaluation of disconnected quark loops is of paramount importance
in order to eliminate a systematic error inherent in the determination
of hadron matrix elements in lattice QCD. For flavor singlet
quantities, these contributions, even though smaller in magnitude as
compared to the connected contributions that are computationally
easier to evaluate, are substantial and cannot be neglected. The
explanation of why these quark loop contributions are large for flavor
singlet quantities is the fact that, in a flavor singlet, the
disconnected contributions coming from different flavors add up, and
hence there is no {\it a priori} reason to neglect them.  Naive
perturbative calculations of some of these flavor singlet
contributions differ from their experimental value, which suggests
that flavor singlet phenomena are inherently linked with
non-perturbative properties of the vacuum. A good example to support
this point is the axial anomaly in the case of the $\eta'$ mass, which
is connected to the topological properties and non-perturbative nature
of QCD.

The computation of disconnected quark loops within the lattice QCD
formulation requires the calculation of all-to-all or time-slice-to-all
propagators, which are impractical to compute exactly, and for which
the computational resources required to estimate them with,
e.g. stochastic methods, are much larger than those required for the
corresponding connected contributions. Therefore, in most hadron
studies up to now the disconnected contributions were neglected
introducing an uncontrolled systematic uncertainty.

Recent progress in algorithms, however, combined with the increase in
computational power, have made such calculations feasible. On the
algorithmic side, a number of improvements like the one-end trick
\cite{Boucaud:2008xu, Michael:2007vn, Dinter:2012tt}, dilution
\cite{Bernardson:1993he, Viehoff:1997wi, O'Cais:2004ww, Foley:2005ac,
  Alexandrou:2012zz}, the Truncated Solver Method (TSM)
\cite{Alexandrou:2012zz, Collins:2007mh, Bali:2009hu} and the Hopping
Parameter Expansion (HPE) \cite{Boucaud:2008xu, McNeile:2000xx} have
led to a significant reduction in both stochastic and gauge noise
associated with the evaluation of disconnected quark loops. Moreover, using special
properties of the twisted mass fermion Lagrangian, one can
further enhance the signal-to-noise ratio by taking the appropriate
combination of flavors. On the hardware side, graphics cards (GPUs) can provide a large speed-up in the evaluation of quark
propagators and contractions. In particular, for the TSM, which
relies on a large number of inversions of the Dirac matrix in single
or half precision, GPUs provide an optimal platform.

In this paper, our aim is to assess recently developed methods and
examine how reliably one can compute disconnected contributions to
flavor singlet quantities by combining the algorithmic advances with
the numerical power of GPUs. We will describe the various improvements
using one ensemble of twisted mass fermion (TMF) gauge field
configurations.  The ensemble is generated with two light degenerate
quarks and a strange and charm quark with masses fixed to their
physical values, referred to as $N_f=2+1+1$ simulations. The lattice
size is $32^3\times 64$, the lattice spacing extracted from the
nucleon mass~\cite{Alexandrou:2013joa} $a=0.082(1)(4)$ and pion mass
about 370~MeV. This ensemble will be hereafter referred to as the
B55.32 ensemble. This paper intends to describe the methodology and
identify the efficiency of the various methods with respect to the
observable under investigation, rather than to arrive at precise
physical results. The latter we reserve for a follow-up
publication. Although we will use the nucleon to test  our
methodology the conclusions apply to any hadron.
The paper is organized as follows: in
Section~\ref{sec:disc_methods} we present the algorithms and variance
reduction techniques we will employ.  In Section~\ref{sec:simulations}
we explain our particular formulation, including information on the
gauge configurations used, as well as details on the GPU
implementation of our methods. Section~\ref{sec:summation} explains
our analysis to extract the desired matrix elements, followed by
Section~\ref{sec:compare} in which we summarize the comparisons
between the different methods employed. In
Section~\ref{sec:conclusions} we give our conclusions and outlook.

\section{Methods for disconnected calculations}
\label{sec:disc_methods}
\subsection{Stochastic estimate\label{stoSec}}
The exact computation of all-to-all (time-slice-to-all) propagators on
a lattice volume of physical interest is outside our current computer
power, since this would require volume (spatial volume) times
inversions of the Dirac matrix, whose size ranges from $\sim 10^7$ for
a $24^3\times 48$ lattice to $\sim 10^9$ for the largest volumes of
$96^3\times 192$ considered nowadays. 
The typical way around this problem is to compute an
unbiased stochastic estimate of the all-to-all propagator
\cite{Bitar:1989dn}. The method consists of generating a set of $N_r$
sources $\left|\eta_r\right\rangle$ randomly, by filling each
component of the source with random numbers drawn from a particular
representation of the $\mathbb{Z}_2$ or $\mathbb{Z}_4$ groups (more
exactly $\left\{1,-1\right\}$ for $\mathbb{Z}_2$ and
$\left\{1,i,-1,-i\right\}$ for $\mathbb{Z}_4$), or from a
representation of $\mathbb{Z}_2\otimes i\mathbb{Z}_2$. Other noise
sets may be used, however it has been shown that $\mathbb{Z}_N$-noise
has smaller variance than e.g. gaussian noise~\cite{Dong:1993pk}.  The
$\mathbb{Z}_N$-noise sources have the following properties:

\begin{eqnarray}	
\frac{1}{N_r}\sum_{r=1}^{N_r}\left|\eta_r\right\rangle = \left|0\right\rangle + \mathcal{O}\left(\frac{1}{\sqrt{N_r}}\right),
\label{eq1} \\
\frac{1}{N_r}\sum^{N_r}_{r=1}\left|\eta_r\right\rangle\left\langle\eta_r\right| = \mathbb{I} + \mathcal{O}\left(\frac{1}{\sqrt{N_r}}\right).
\label{eq2}
\end{eqnarray}
The first property ensures that our estimate of the propagator is unbiased. The second one allows us to reconstruct the inverse matrix by solving for $\left|s_r\right\rangle$ in
\begin{equation}
M\left|s_r\right\rangle = \left|\eta_r\right\rangle
\label{etaToS}
\end{equation}
and calculating
\begin{equation}
M_E^{-1}:=\frac{1}{N_r}\sum_{r=1}^{N_r}\left|s_r\right\rangle\left\langle\eta_r\right|\approx M^{-1}.
\label{estiM}
\end{equation}
\noindent Since in general the number of noise vectors $N_r$ required
is much smaller than the lattice volume $V$, the computation
becomes feasible, although it can still be very expensive depending on
the value of $N_r$ required to achieve a good estimate of $M^{-1}$ in
Eq. \eqref{estiM}.

The deviation of our estimator from the exact solution is given by
\begin{equation}
M^{-1}-M_E^{-1} = M^{-1}\times\left(\mathbb{I}-\frac{1}{N_r}\sum_{r=1}^{N_r}\left|\eta\right\rangle\left\langle\eta\right|\right),
\label{errS}
\end{equation}
so as $N_r$ increases the introduced stochastic error decreases, as
Eq. \eqref{eq2} clearly shows. In fact, from Eqs. \eqref{eq2},
\eqref{errS} we see that the errors decrease as
$\mathcal{O}\left(\frac{1}{\sqrt{N_r}}\right)$, as expected from the
properties of these noise sources.

Since we have to deal with gauge error, i.e. the error coming from the
fact that we analyze a representative set of gauge configurations, the
number of stochastic noise sources should be taken so that the stochastic
error is comparable to the gauge error. This criterion ideally determines
the number of stochastic sources $N_r$, which can  differ for
each observable. Since we will be interested in evaluating a range of
observables we will choose $N_r$ that can yield good results for the 
most demanding among these observables.

\subsection{The Truncated Solver Method\label{secTSM}}

The Truncated Solver Method (TSM)~\cite{Alexandrou:2012zz,
  Collins:2007mh, Bali:2009hu} is a way to increase $N_r$ at a reduced
computational cost. The idea behind the method is the following:
instead of inverting to high precision the stochastic sources in
Eq. \eqref{etaToS}, we can aim at a low precision (LP)
estimate \begin{equation} \left|s_r\right\rangle_{LP} =
  \left(M^{-1}\right)_{LP}\left|\eta_r\right\rangle,
\label{etaToSLP}
\end{equation}
where the inverter, which is a Conjugate Gradient (CG) solver in this
work, is truncated. The truncation criterion can be a low precision
stop condition for the residual (for instance, $|\hat{r}|< 10^{-2}$,
with $\hat{r}$ the residual vector in the CG algorithm), or a fixed
number of iterations, roughly around $1/10$ or $1/20$ of what would
be needed to obtain a high precision (HP) solution. This way we can
increase the number of stochastic sources $N_{\rm LP}$ at a very small
cost.  Using the low precision sources our estimate of the inverse
matrix given by Eq. \eqref{estiM} is not unbiased, so we are
introducing new errors in the computation of the all-to-all
propagator.

In order to correct for the bias introduced using low precision, we estimate the correction $C_E$ to this bias stochastically by inverting a number of sources to high and low precision, and calculating the difference,
\begin{equation}
C_E:=\frac{1}{N_{\rm HP}}\sum_{r=1}^{N_{\rm HP}}\left[\left|s_r\right\rangle_{\rm HP} - \left|s_r\right\rangle_{\rm LP}\right]\left\langle\eta_r\right|,
\label{estiC}
\end{equation}
where the $\left|s_r\right\rangle_{\rm HP}$ are calculated by solving Eq. \eqref{etaToS} up to high precision, so our final estimate becomes

\begin{eqnarray}
M_{E_{TSM}}^{-1}:=& &\frac{1}{N_{\rm HP}}\sum_{r=1}^{N_{\rm HP}}\left[\left|s_r\right\rangle_{\rm HP} - \left|s_r\right\rangle_{\rm LP}\right]\left\langle\eta_r\right| \nonumber \\
&+&
\frac{1}{N_{\rm LP}}\sum_{j=N_{\rm HP}+1}^{N_{\rm HP}+N_{\rm LP}}\left|s_r\right\rangle_{\rm LP}\left\langle\eta_r\right|,
\label{estiTSM}
\end{eqnarray}
which requires $N_{\rm HP}$ high precision inversions and $N_{\rm
  HP}+N_{\rm LP}$ low precision inversions. Following the discussion
in Ref.~\cite{Blum:2012uh}, one expects the error of this
improved estimate of the fermion loop to scale as:
\begin{equation}
e\sqrt{2(1-r_c)+\frac{1}{n_{\rm LP}}},
\end{equation}
 where the unimproved error $e$ scales as $\propto 1/\sqrt{N_{\rm
     HP}}$ and $n_{\rm LP}=N_{\rm LP}/N_{\rm HP}$. 
 $r_c$ is the correlation between the
 $N_{\rm HP}$ quark propagators in low and high precision, which
 is expected to be close to unity (with the optimal being one)
 and depends on the criterion for the LP
 inversions and on how well-conditioned the Dirac fermion matrix is. 
In this work, we use the twisted mass formulation for the fermion
action, hence the smallest eigenvalues depend on the value of the
twisted mass parameter $\mu$, and our matrix is protected from
near-zero eigenvalues.

In the TSM one needs to tune the precision of the LP inversions as
well as the $n_{\rm LP}$ ratio, with the goal of choosing as large a
ratio as possible while still ensuring that the final result is
unbiased and that $r_c\simeq 1$. In the next subsection we give details
on how we optimized the TSM parameters with this criterion in mind.

\subsubsection{Tuning the TSM parameters}
In order to achieve good performance for the TSM there are two
parameters to be tuned, namely the number of noise vectors $N_{\rm
  LP}$ computed with low precision and the number of noise vectors
$N_{\rm HP}$ computed at  high precision. The
criterion for the low precision inversions can be selected by
specifying a relaxed stopping condition in the conjugate gradient,
e.g. by allowing a relatively large value of the residual, which will
in turn determine the number of iterations required to invert a LP
source. Following Ref.~\cite{Alexandrou:2012zz}, we choose a stopping
condition at fixed value of the residual $|\hat{r}|_{\rm LP}\sim
10^{-2}$. $N_{\rm HP}$ is selected by requiring that the bias
introduced when using $N_{\rm LP}$ low precision vectors is corrected.
Our goal is to develop methods for computing fermion loops with the
complete set of $\Gamma$-matrices up to one-derivative operators.
The tuning is, thus, performed using an operator that requires a large
number of stochastic noise vectors, such as $g_A$ or equivalently the
nucleon momentum fraction $\langle x\rangle$ and we optimized $N_{\rm
  HP}$ and $N_{\rm LP}$ so as to get the smallest error at the lowest
computational cost.

\begin{figure}[h!]
\includegraphics[width=0.4\textwidth,angle=0]{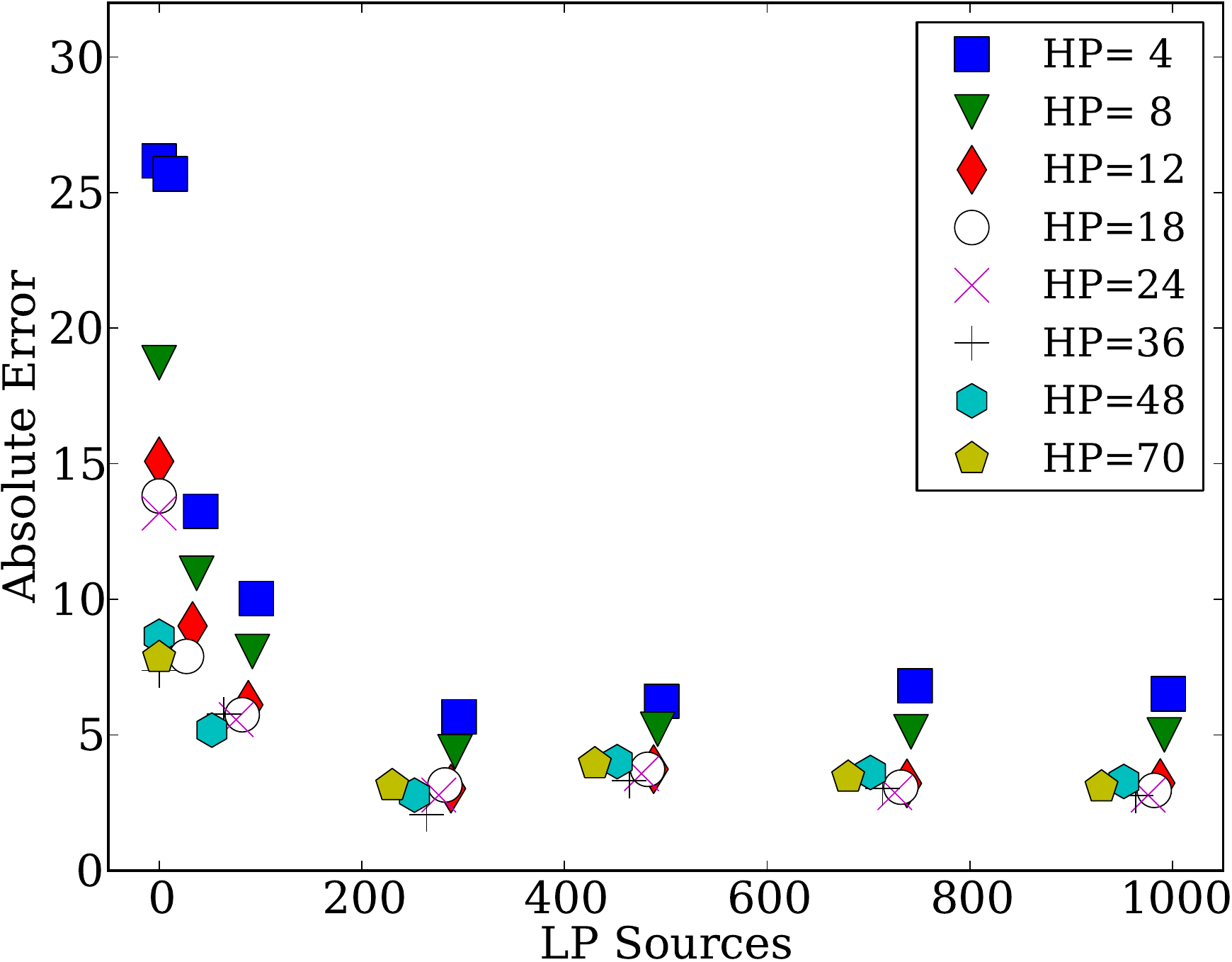}
\caption{Tuning of $N_{\rm HP}$ and $N_{\rm LP}$ entering the TSM using the  B55.32 ensemble on 50 configurations for the
nucleon matrix element of the 
  operator $i\bar\psi\gamma_3 D_3\psi$.
The insertion time  is fixed at  $t_{\rm ins}= 8a$ and sink time
at  $t_{\rm s} = 16a$. The error is shown versus $N_{\rm LP}$ for
  different values of $N_{\rm HP}$ marked by the different plotting
  symbols as indicated in the legend.
\label{TSMPerfPlot}}
\end{figure}

\begin{figure}[h!]

\includegraphics[width=0.42\textwidth,angle=0]{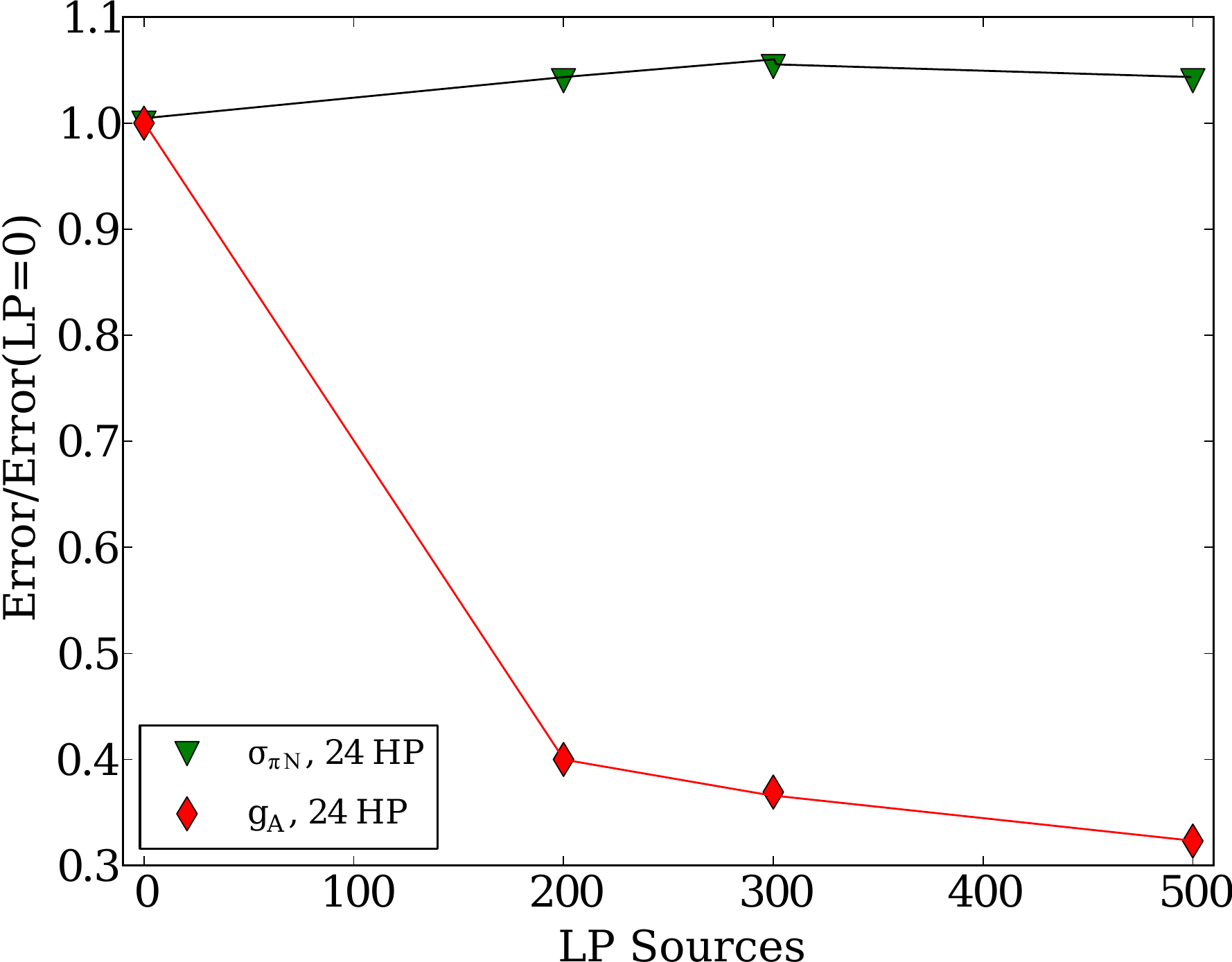}
\caption{The error versus $N_{\rm LP}$ fixing $N_{\rm HP}=24$ for 
 $\sigma_{\pi N}$ and the isoscalar $g_A$ for 56400
  measurements.\label{compTSMs}\label{TSMConvPlot}}

\end{figure}

In Fig.~\ref{TSMPerfPlot} we show the error on the nucleon
matrix element of the vector one-derivative operator related to the  momentum fraction
as a function of $N_{\rm LP}$
for different $N_{\rm HP}$. For
$N_{\rm LP}=0$, we observe that the error decreases as the number of
HP noise vectors increases, as expected, but saturates when $N_{\rm HP}=36$. 
For $N_{\rm LP}\ne0$ we see that the error saturates for
$N_{\rm LP} \stackrel{>}{\sim} 200$ for this small test ensemble of 50 configurations,
and no further improvement is observed as $N_{\rm LP}$
increases. For $N_{\rm LP}=200-300$ 
we observe that we need at least $N_{\rm HP}=8-12$ to correct the bias or
$n_{\rm LP}\sim 20$.
The value of the optimal ratio
 $n_{\rm LP}$ needed for different loops varies depending on the 
observable. This is demonstrated in Fig.~\ref{TSMConvPlot} where we show the
relative error in the case of the isoscalar axial charge $g_A$ and the light
quark $\sigma$-term, $\sigma_{\pi N}=\frac{m_u+m_d}{2}\langle N|\bar{u}u+\bar{d}d|N\rangle$ for $N_{\rm HP}=24$.  As can be seen,
in the case of $g_A$ one requires at least  $N_{\rm LP}=500$,  while for the
$\sigma_{\pi N}$-term  $N_{\rm LP}=0$ is
sufficient making the TSM unnecessary.

\begin{figure}[h!]

\includegraphics[width=0.42\textwidth,angle=0]{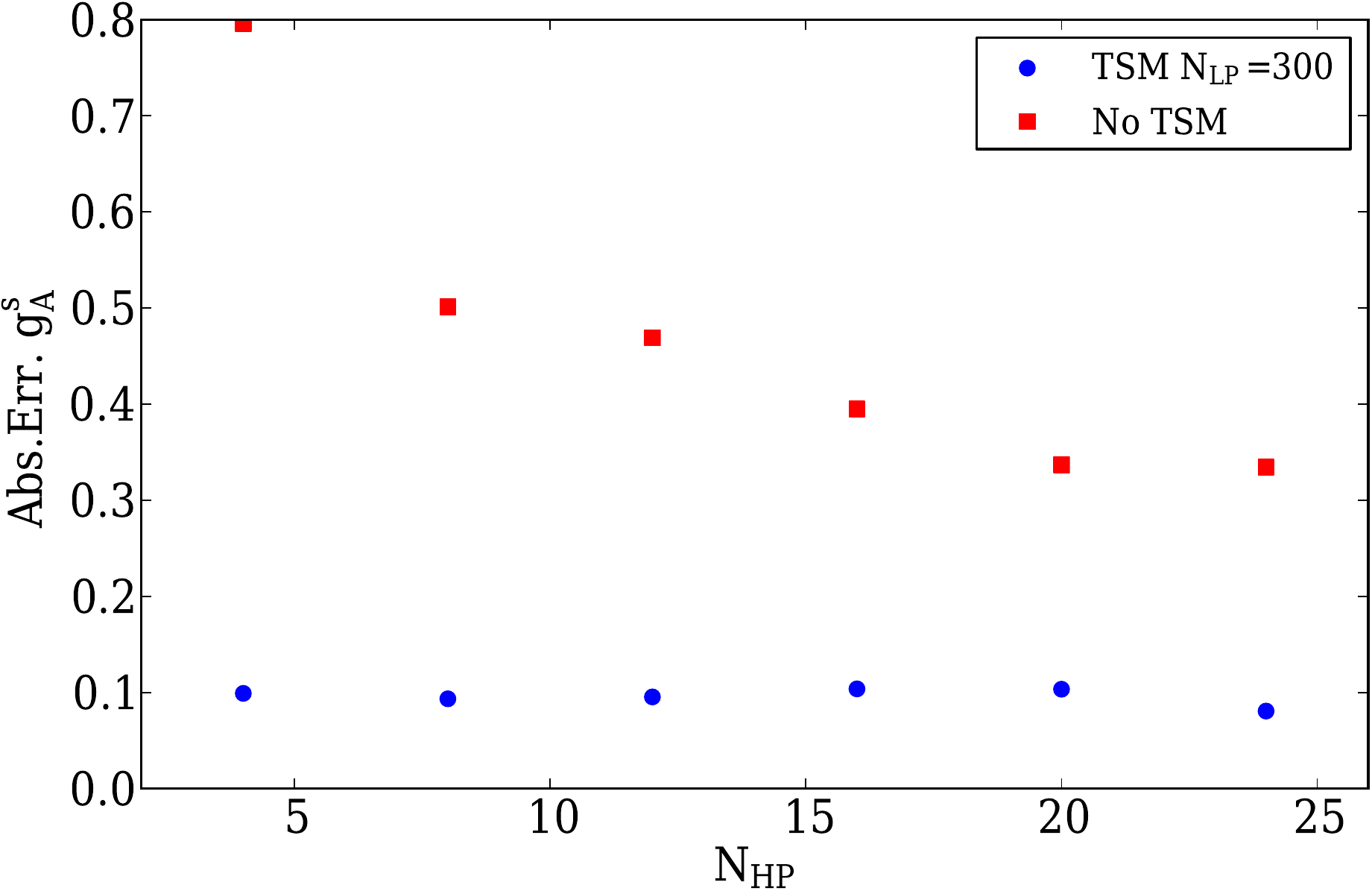}\\
\includegraphics[width=0.42\textwidth,angle=0]{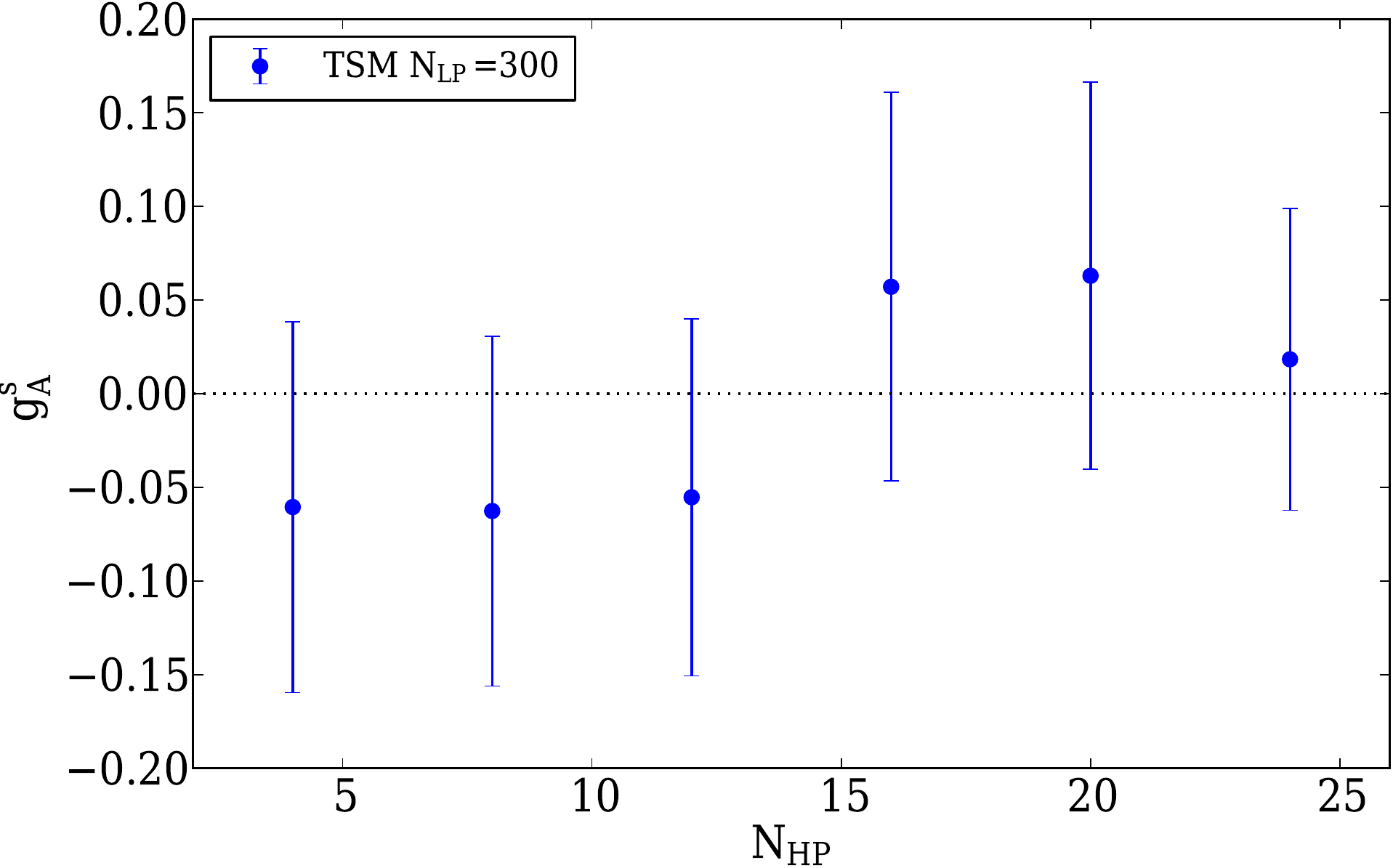}
\caption{The error (upper) and the mean value (lower) versus $N_{\rm HP}$ fixing $N_{\rm LP}=300$ for 
 $g_A^s$ using 448 configurations.
 }
\label{fig:convergence}

\end{figure}

In assessing the various methods we will be using $N_{\rm HP}=8$ and $N_{\rm LP}=200-300$.
In  Fig.~\ref{fig:convergence} we show for $N_{\rm LP}=300$ the error on the
strange quark loop contribution to the nucleon axial charge $g_A^s$ as a function of $N_{\rm HP}$. As can be seen, the error remains constant as 
 $N_{\rm HP}$ is increased from 4 to 24. In addition, we show the
mean value, which  is also consistent for different $N_{\rm HP}$ within the current statistics. 
Therefore choosing $N_{\rm HP}=8$ 
and $N_{\rm LP}\le 300$ does not introduce any bias on the error.

\subsection{The one-end trick}

The twisted mass fermion (TMF) formulation allows the use of a very powerful
method to reduce the variance of the stochastic estimate of the
disconnected diagrams. From the discussion given in
section~\ref{stoSec}, the standard way to proceed with the computation
of disconnected diagrams would be to generate $N_r$ stochastic sources
$\eta_r$, invert them as indicated in Eq.~\eqref{etaToS}, and compute
the disconnected diagram corresponding to an operator $X$ as

\begin{eqnarray}
\frac{1}{N_r}\sum_{r=1}^{N_r} \left\langle \eta^\dagger_r X s_r\right\rangle &=& \textrm{Tr}\left(M^{-1}X\right) \nonumber \\ & +& O\left(\frac{1}{\sqrt{N_r}}\right),
\label{loopSt}
\end{eqnarray}
where the operator $X$ is expressed in the twisted basis. However, if the operator $X$ involves a $\tau_3$ acting in flavor space, one can utilize the following identity of the twisted mass Dirac operator with $+\mu$ denoted by $M_u$ and $-\mu$ denoted by $M_d$: 

\begin{equation}
  M_u - M_d = 2i\mu a\gamma_5.
\end{equation}
Inverting this equation we obtain

\begin{equation}
M^{-1}_u - M^{-1}_d = -2i\mu aM_d^{-1}\gamma_5M_u^{-1}.
\label{vvTrick}
\end{equation}
Therefore, instead of using Eq. \eqref{loopSt} for the operator $X\tau_3$, we can alternatively write

\begin{align}
\frac{2i\mu a}{N_r}\sum_{r=1}^{N_r} \left\langle s^\dagger_r \gamma_5 X s_r\right\rangle =& \nonumber\\
\textrm{Tr}\left(M_u^{-1}X\right) - \textrm{Tr}\left(M_d^{-1}X\right) &+ O\left(\frac{1}{\sqrt{N_r}}\right)=\nonumber\\
-2i\mu a\textrm{Tr}\left(M_d^{-1}\gamma_5M_u^{-1}X\right) &+ O\left(\frac{1}{\sqrt{N_r}}\right).
\label{loopVv}
\end{align}

Two main advantages emerge due to this substitution: i) the fluctuations are effectively reduced by the $\mu$ factor, which is small in current simulations, and
ii)  an
implicit sum of $V$ terms  appears in the right hand side (rhs) of Eq.~\eqref{vvTrick}. The trace of the left hand side (lhs) of the same equation develops a signal-to-noise ratio of $1/\sqrt{V}$, but thanks to
this implicit sum, the signal-to-noise ratio of the rhs becomes $V/\sqrt{V^2}$. In fact, using the
one-end trick yields for the same operator  a large reduction in the errors for the same computational
cost as compared to not using it~\cite{Boucaud:2008xu, Michael:2007vn, Dinter:2012tt}.
The  identity given in Eq.~\ref{vvTrick} can only be applied when a $\tau_3$ flavor matrix appears in the operator expressed in the twisted basis. For other operators one 
can  use the identity

\begin{equation}
M_u + M_d = 2D_W,
\end{equation}
where $D_W$ is the Dirac-Wilson operator without a twisted mass term. After some algebra, one finds

\begin{eqnarray}
\frac{2}{N_r}\sum_{r=1}^{N_r} \left\langle s^\dagger_r \gamma_5 X\gamma_5 D_W s_r\right\rangle &=& \textrm{Tr}\left(M_u^{-1}X\right) + \textrm{Tr}\left(M_d^{-1}X\right) \nonumber \\ & +& O\left(\frac{1}{\sqrt{N_r}}\right).
\label{loopStD}
\end{eqnarray}
Computing the fermion loops in this way, which we will hereafter refer
to as the generalized one-end trick, lacks the $\mu$-suppression
factor, which, as we will see, introduces a considerable penalty in
the signal-to-noise ratio.

Because of the volume sum that appears in Eq.~\eqref{vvTrick} and
Eq.~\eqref{loopStD}, the sources must have entries on all sites, which
in turn means that we can compute the fermion loop at all insertions in a
single inversion. This allows us to evaluate the three-point function for
all combinations of source-sink separation and insertion time-slices,
which will prove essential in identifying the contribution of excited
state effects for the different operators.

\subsection{Time-dilution}

For isovector operators in the twisted mass basis the best approach,
as we will discuss in the next section, is to use the identity given
in Eq.~(\ref{vvTrick}) that takes advantage of the $\mu$-suppression
factor.  For other operators the method of choice is not clear and
different variance reduction techniques may be more efficient than the
generalized one-end trick and need to be considered. One approach that
is used to reduce stochastic noise is dilution, i.e. instead of
filling up all the entries of the source vector, we 
populate only parts  by
decomposing the space
$\mathcal{R}=V\oplus$color$\oplus$spin in $m$ smaller
subspaces given by the direct sum $\mathcal{R}=\sum_{i=1}^m\mathcal{R}_i$, and defining our
noise sources in those subspaces. This way, Eq.~\eqref{estiM} still
holds, but a reduction in the variance of the disconnected diagrams
may result. This expectation can be seen by examining Eq.~\eqref{errS}
where the contributions to the noise come from the off-diagonal terms
of $M^{-1}$, since the matrix
$\mathbb{I}-\frac{1}{N_r}\sum_{r=1}^{N_r}\left|\eta_r\right\rangle\left\langle\eta_r\right|$
features only off-diagonal entries. The off-diagonal terms decrease
exponentially with the source-sink separation, so the neighboring
terms to the sink have the strongest influence on the errors, hence a
dilution in space-time could prove useful in reducing the noise. Noise
can also come from strongly coupled spin components, and dilution in
color has also been shown to be successful in some systems. In the
end, for a given number of subspaces $m$, whenever the reduction of
errors surpasses the factor $1/\sqrt{m}$, dilution becomes
advantageous.  This cost of inversions can be reduced by using
deflated solvers~\cite{Stathopoulos:2007zi,Stathopoulos:2009zz}, which become more efficient as the number of rhs
increases, thereby improving the performance of this approach.

In this work, we examine whether time-dilution can bring an
improvement for the operators where Eq.~(\ref{vvTrick}) can not be
applied.  One can apply the coherent
method~\cite{Alexandrou:2010uk,Syritsyn:2009mx} using noise vectors
with entries in several time slices, as long as these time slices are
far enough from each other, so that only a single loop contributes,
thus increasing the statistics at almost no cost. For operators
involving a time derivative, one would need additional inversions at
time slices $t-1$ and $t+1$ effectively tripling the required
computational time. Therefore, for the current work where we focus on
comparisons of the different methods, we restrict ourselves to
examining ultra-local current insertions, i.e. loops having an
insertion of the form $\overline{\psi}(x)\Gamma\psi(x)$ .

\subsection{Hopping Parameter Expansion}

Another technique that can be used to reduce the variance of our
estimate of the propagators is the \emph{Hopping Parameter Expansion}
(HPE). The idea is to expand the inverse of the fermionic matrix in
terms of the hopping parameter~\cite{Thron:1997iy},

\begin{eqnarray}
M_u^{-1} = B - BHB + \left(BH\right)^2B &-& \left(BH\right)^3B \nonumber \\
 &+& \left(BH\right)^4M_u^{-1},
\label{HPE}
\end{eqnarray}
where $B=\left(1+i2\kappa\mu a\gamma_5\right)^{-1}$ and $H=2\kappa \Dsla$, with $\Dsla$ the hopping term. The first four terms in this expansion can be computed exactly, while the fifth term
is calculated stochastically as

\begin{eqnarray}
\frac{1}{N_r}\sum_{r=1}^{N_r} \left[X\left(BH\right)^4 s_r \eta^\dagger_r\right] &=& \textrm{Tr}\left[X\left(BH\right)^4 M_u^{-1}\right] \nonumber \\
 &+& O\left(\frac{1}{\sqrt{N_r}}\right).
\label{loopHPE1}
\end{eqnarray}

The first term in Eq. \eqref{HPE} is the only one that does not
involve the gauge links, and is non-zero for ultra-local operators whose
$\gamma$-structure is proportional to $I$ or $\gamma_5$. The rest of the
terms include the hopping matrix, which is traceless, so only the even
powers (third term) will survive for ultra-local operators. Moreover, if $X$
is not proportional to $I$ or $\gamma_5$, the third term is zero as
well, since the resulting matrix is traceless. For one-derivative
operators, only the second and fourth terms survive, provided that
$X$ is proportional to either $I$ or $\gamma_5$ (or a linear
combination of the two). In any case, since these terms are computed
in advance and do not depend on the gauge configuration for local
operators, they do not incur a serious computational overhead.

\section{Simulation details}
\label{sec:simulations}
 As already mentioned, we examine the performance of the various
 methods by analyzing an ensemble of $N_f=2+1+1$ twisted mass
 configurations simulated with pion mass of $ am_\pi=0.15518(21)(33)$
 and strange and charm quark masses fixed to approximately their
 physical values (B55.32 ensemble)~\cite{Baron:2010bv}.  The lattice
 size is $32^3\times 64$ giving $m_{\pi}L\sim 5$.

For the disconnected diagrams we make use of a
modified version of the QUDA library~\cite{Clark:2009wm,Babich:2011np},
 in which we implement new host code   and
CUDA kernels to enable the required inversions and contractions on NVIDIA
GPUs. In particular, we developed a CUDA version for the twisted-mass fermion operator with even-odd
preconditioning to allow QUDA inverters to work with this regularization.
We also developed new kernels that carry out  efficiently the contractions
of the quark propagator and the  current insertion, as well as all the interfaces
required to make use of these new additions to QUDA. Whereas the new 
twisted-mass operator is available in the official QUDA release, the contraction
kernels are currently available in the local branch of the QUDA git repository,
\footnote{https://github.com/lattice/quda/tree/discLoop}. 
Details on the implementation of the twisted-mass dslash
operator and the new contraction kernels can be found in  Appendix \ref{sec:app1}.

\begin{figure}[h!]

\includegraphics[width=0.42\textwidth,angle=0]{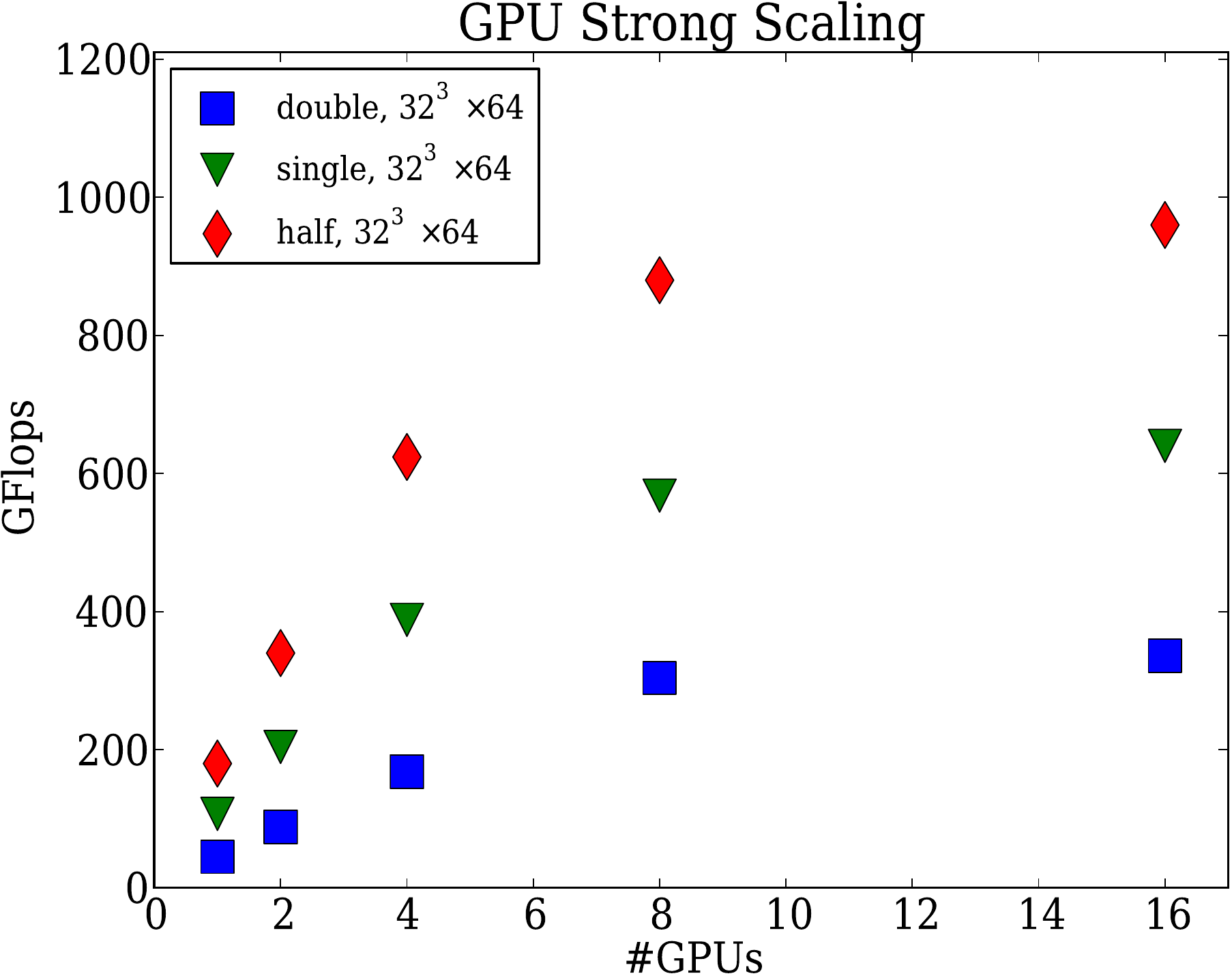}
\caption{Strong scaling of the multi-GPU conjugate-gradient solver using the B55.32
  ensemble and either 64-bit (double), 32-bit (single) or 16-bit (half) floating
  point precision.\label{PerfPlots2}}

\end{figure}

\begin{figure}[h!]

\includegraphics[width=0.42\textwidth,angle=0]{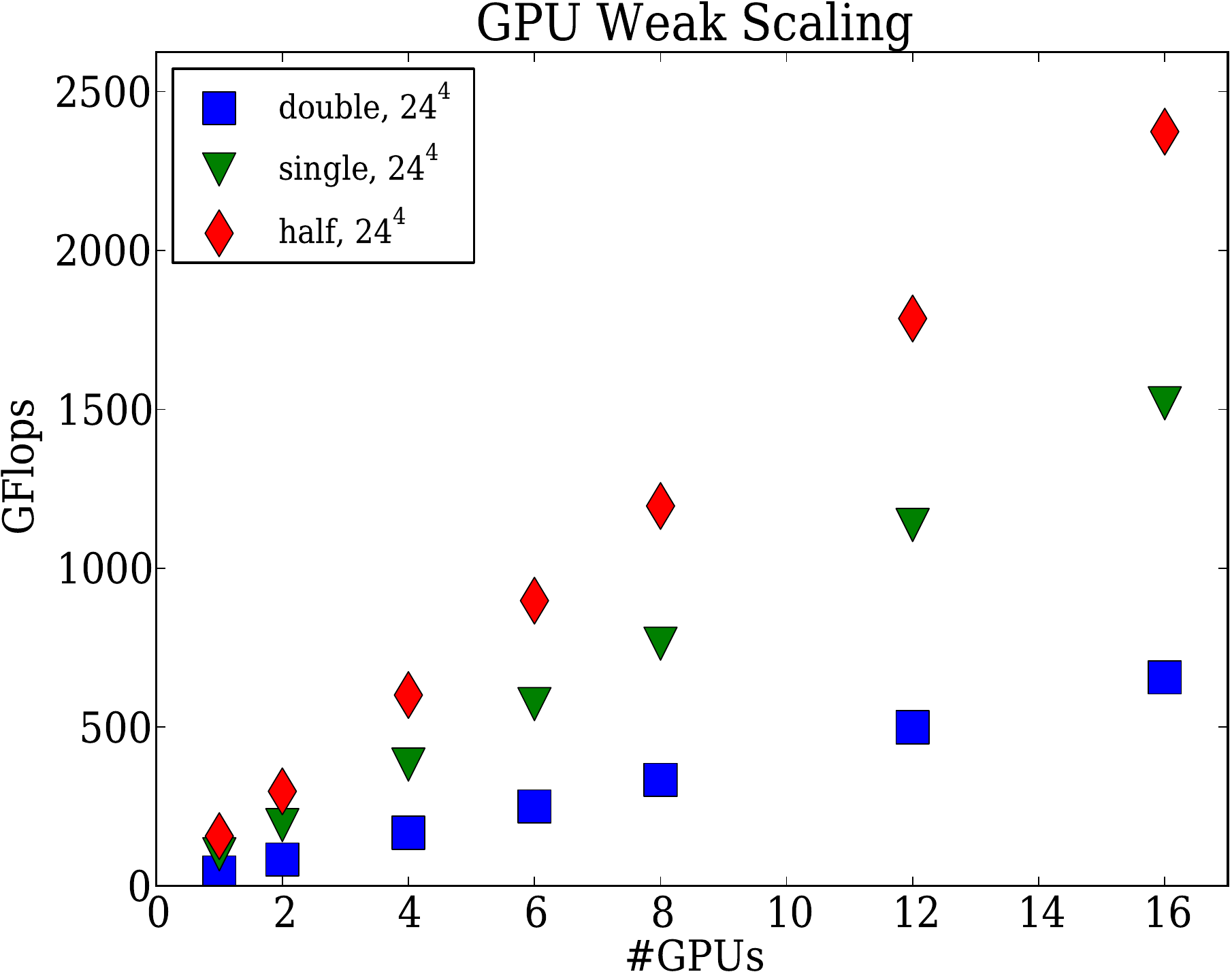}
\caption{Weak scaling of the multi-GPU conjugate-gradient solver
   for a local volume
  $V=24^4$, using the same notation as in Fig.~\ref{PerfPlots2}\label{PerfPlots}}

\end{figure}

For the Fourier transform we use the CUFFT library.  The
QUDA library allows for inter- and intra-node multi-GPU calculations through MPI;
since in our setup there are two GPUs per node, and the GPU-memory requirements for
the contractions are high, we use 2 GPUs working in parallel by splitting the lattice
between them. In Figs.~\ref{PerfPlots2} and \ref{PerfPlots} we show strong and weak
scaling as a function of the number of GPUs. Strong scaling is good for a few GPUs,
with a $\sim90\%$ increase in performance when adding the second GPU. This result
holds for up to 8 GPUs in the strong scaling case, after which we observe a drop in
performance. For the architecture on which we carried out these calculations, namely
dual M2070 NVIDIA GPU equipped nodes over a QDR infiniband, the only advantage in going
beyond 8 GPUs seems to be in the case where GPU memory is insufficient. As can be seen,
we can reach TFlop sustained performance with just a few GPUs. Weak scaling on the
other hand is almost perfect, which can be understood if one considers that GPUs perform
optimally the larger the local lattice size.


The noise sources are generated on-the-fly, and the propagators are not stored,
in order to save storage and I/O time.

\section{Analysis with the plateau and summation methods}
\label{sec:summation}
One of the advantages of the one-end trick for twisted mass fermions
is the fact that, since the noise sources are defined on all sites, we
obtain the fermion loops at all insertion time-slices. We can thus
compute all possible combinations of source-sink separations and
insertion times in the three-point function. This feature enables us
to use the summation method, in addition to the plateau method, with
no extra computational effort.

The summation method has been known since a long
time~\cite{Maiani:1987by, Gusken:1999te} and has been revisited in the
study of $g_A$~\cite{Capitani:2010sg}. In both the plateau and
summation approaches, one constructs ratios of three- to two-point
functions in order to cancel unknown overlaps and exponentials in the
leading contribution such that the matrix element of the ground state
is isolated. For zero-momentum transfer we consider the ratio \be
R(t_{ins},t_s)=\frac{G^{3pt}(t_{ins},t_s)}{G^{2pt}(t_s)}\>,
\label{ratio}
\ee where $G^{3pt}(t_{ins},t_s)$ and $G^{2pt}(t_s)$ are the three- and
two-point functions at zero momentum, respectively. The leading time
dependence of this ratio is given by
\begin{equation}
R(t_{ins},t_s) = R_{GS} + O(e^{-\Delta E_K t_{ins}}) + O(e^{-\Delta E_K(t_s-t_{ins})}),
\label{RatioPlaDet}
\end{equation}
where $R_{GS}$ is the matrix element of interest, and the other
contributions come from the undesired excited states of energy
difference $\Delta E_K$. In the plateau method, one plots
$R(t_{ins},t_s)$ as a function of $t_{ins}$, which should be a
constant (plateau region) when excited state effects are negligible.
A fit to a constant in the plateau region thus yields  $R_{GS}$. In the alternative
summation method, one performs a sum over $t_{ins}$ to obtain:

\begin{equation}
R_{\rm sum}(t_s) = \sum_{t_{ins}=0}^{t_{\rm ins}=t_s} R(t_{\rm ins},t_s) = t_sR_{GS} + a + O(e^{-\Delta E_Kt_s})
\label{RatioSumDet}
\end{equation}
and now the exponential contributions coming from the excited states
decay as $e^{-\Delta E_K t_s}$ as opposed to the plateau method where
excited states are suppressed like $e^{-\Delta E_K (t_s-t_{ins})}$,
with $0\le t_{\rm ins}\le t_s$, the insertion time. Therefore, we
expect a better suppression of the excited states for the same $t_s$.
Note that one can exclude from the summation $t_{\rm ins}=t_s$ and
$t_{\rm ins}=0$ without affecting the dependence on $t_s$ in
Eq.~\eqref{RatioSumDet}. The results given in this work are obtained
excluding these contact terms from the summation.  The drawback of the
summation method is that one requires knowledge of the three-point
function for all insertion times and that we need to fit to a straight
line with two fitting parameters instead of one.

\section{Comparison of results of different methods}
\label{sec:compare}

\begin{figure}[h!]

\includegraphics[width=0.42\textwidth,angle=0]{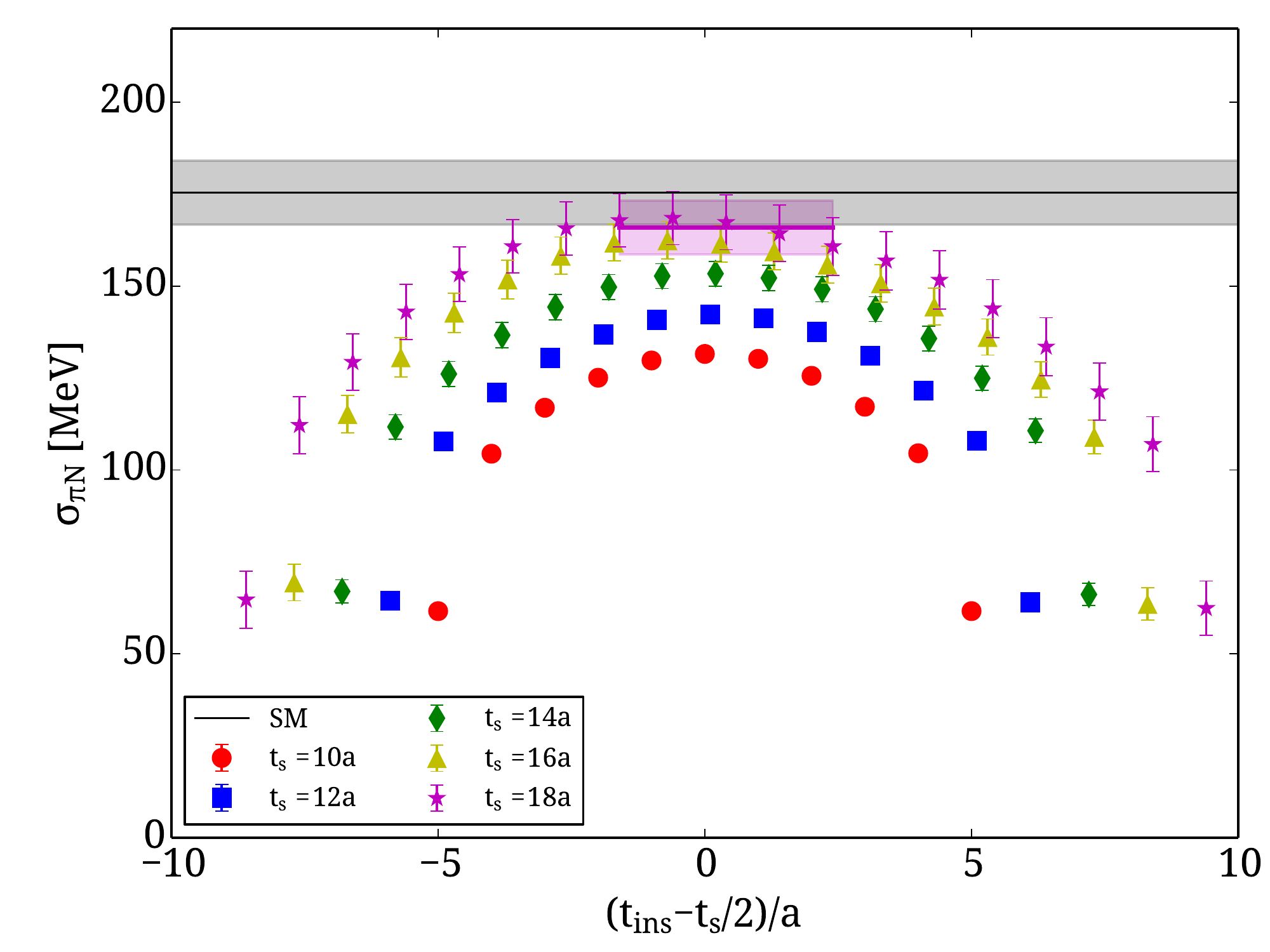}\\
\includegraphics[width=0.42\textwidth,angle=0]{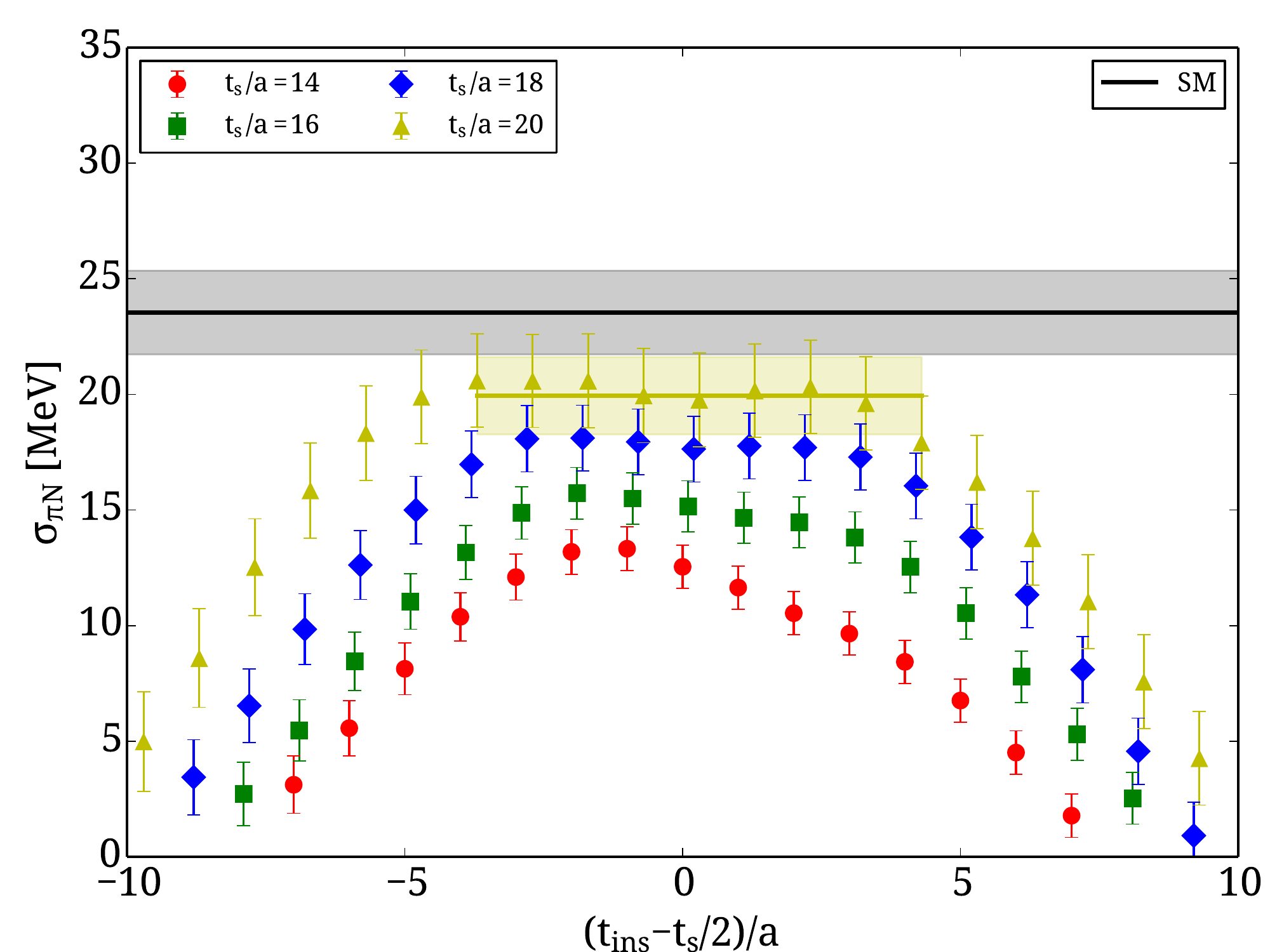}
\caption{Excited states contribution to the
connected (upper) and disconnected (lower) ratios 
for $\sigma_{\pi N}$.  The 
        ratio is shown as a function of the insertion time-slice with respect
        to mid-time separation ($t_{\rm ins}-t_s/2$) for various source-sink
        separations, $t_{\rm s}$.  The gray band is the
         result obtained from the summation method while the colored band
         is the result of the constant fit in the plateau region.}

\label{fig:excited sigma}
\end{figure}

In order to compare the various methods, we focus on two quantities
with very different behaviors: the $\sigma$--term, for which the
stochastic noise is small and thus a relatively small number of noise
sources are required, and $g_A$ that belongs to a class of observables
which require a large number of noise vectors and statistics to be
computed in a reliable way.  These two quantities are also different
with respect to excited states contamination, with the $\sigma$-terms
having large excited state
contributions~\cite{Alexandrou:2012gz,Dinter:2012qt} while $g_A$ was
shown to be less affected~\cite{Dinter:2011sg,Alexandrou:2011aa},
although the degree of contamination may depend on the value of the
pion mass~\cite{Green:2012ud,Green:2012rr,Capitani:2012gj}. We note in
particular that the summation method as applied in the extraction of
$g_A$ in Ref.~\cite{Capitani:2012gj} led to agreement with the
physical value after performing a chiral extrapolation, while in
Ref.~\cite{Green:2012ud} it was shown that the value extracted using
the summation method at near physical pion mass is reduced further
away from the physical value, possibly due to thermal
effects~\cite{Green:2012rr}. On the other hand, a high statistics
analysis for the ensemble used in this work showed no excited states
contamination for $g_A$~\cite{Dinter:2011sg}, while for the
$\sigma-$term for the same ensemble we find large contributions from
excited states. We expect the excited states contribution to
behave similarly for the connected and disconnected three-point functions.
This has been verified in the case of the nucleon $\sigma_{\pi N}$ as shown in Fig.~\ref{fig:excited sigma} where we show both the connected and disconnected
contributions.

In this work we evaluate the light disconnected contributions, the
strange and charm quark contributions to both of these observables
with the one-end trick. In addition, we calculate the strange quark
contribution when using time-dilution, both with and without the HPE
and compare the results.  Regarding the renormalization of the
$\sigma$-terms, the twisted mass formulation has the additional
advantage of avoiding any mixing, even though we are using Wilson-type
fermions~\cite{Dinter:2012tt}.  For the case of the axial charge,
renormalization involves mixing from the three quark sectors. For the
tree-level Symanzik improved gauge action this mixing was shown to be
a small effect of a few percent~\cite{Skouroupathis:2008mf}. We expect
this to hold also for the Iwasaki action used in this work. Since the
main goal of this paper is the comparison of methods, we renormalize
the axial charge neglecting any mixing using the same renormalization
constant as the purely connected part, that is, by multiplying by
$Z_A$.

\subsection{Efficiency of TSM}

  \begin{figure*}[h!]
    
      \begin{minipage}{0.325\linewidth}
		\includegraphics[width=\linewidth,angle=0]{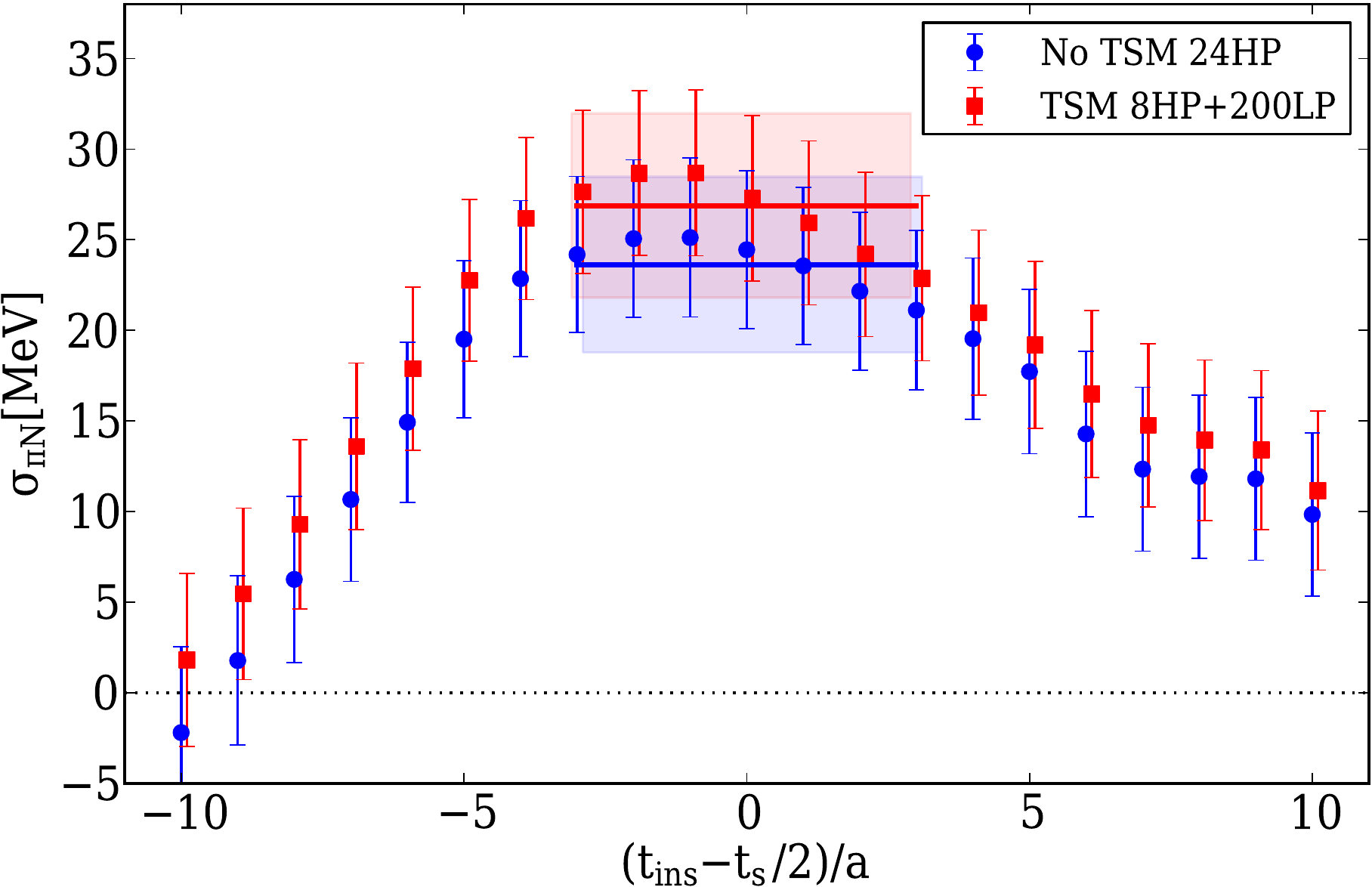}
      \end{minipage}
      \begin{minipage}{0.325\linewidth}
        \includegraphics[width=\linewidth,angle=0]{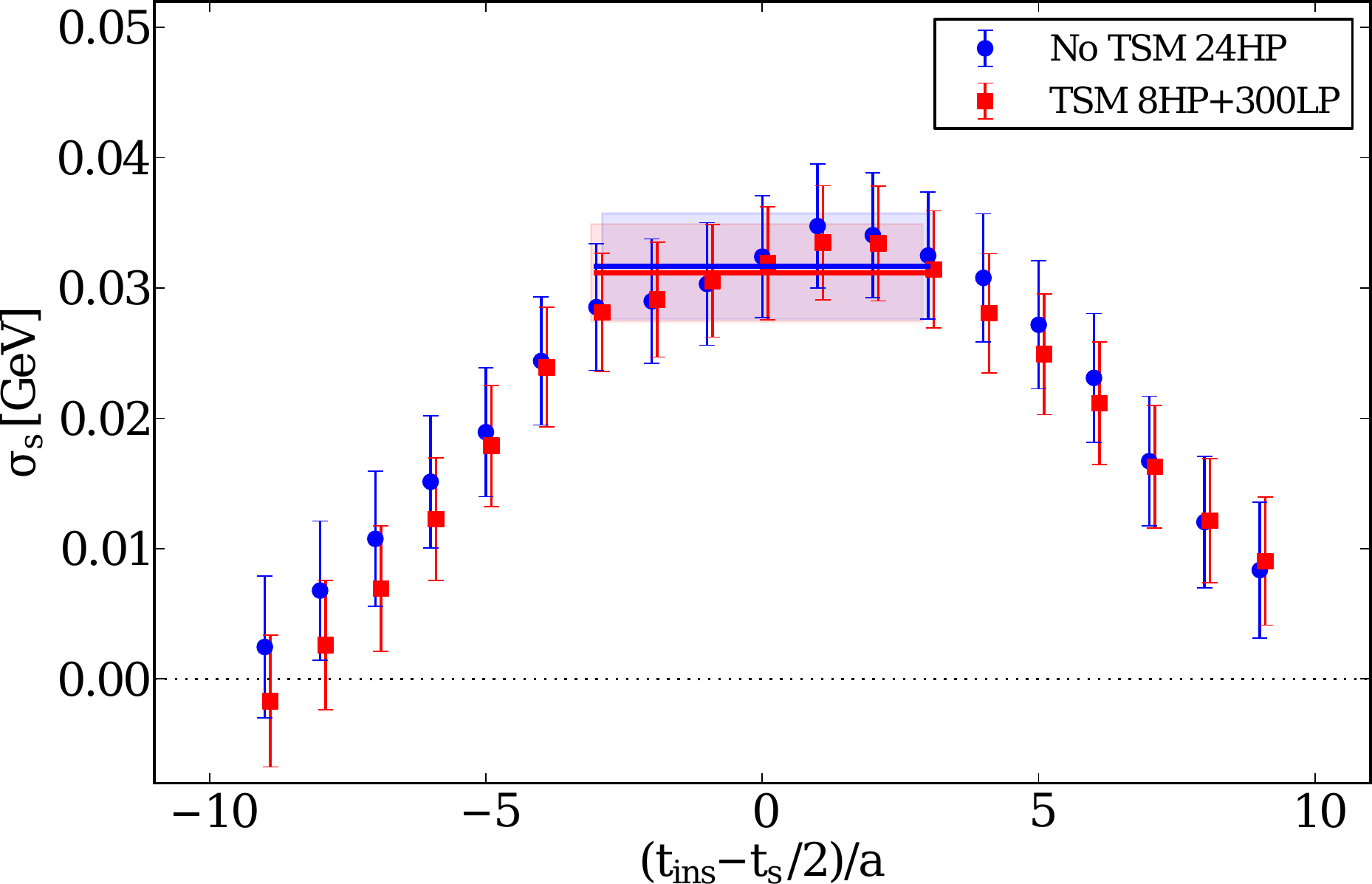} 
      \end{minipage}
      \begin{minipage}{0.325\linewidth}
        \includegraphics[width=\linewidth,angle=0]{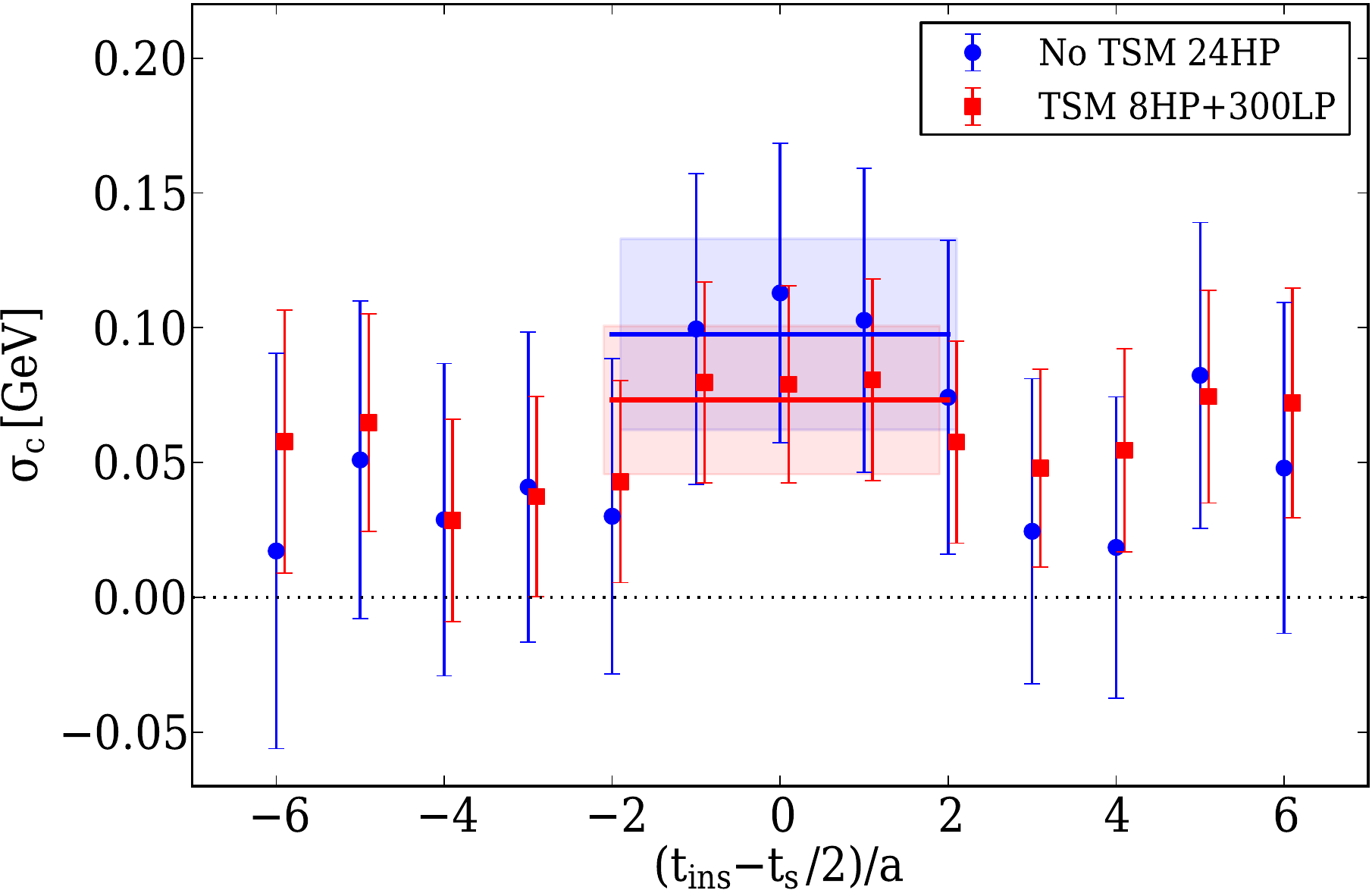} 
      \end{minipage}
	  \caption{Comparison of results obtained using $N_{\rm
              HP}=24$ with $N_{\rm LP}=0$ (no TSM) with those obtained
            using the TSM for $N_{\rm HP}=8$ and $N_{\rm LP}=300$,
            except for the light sector, which uses $N_{\rm
              LP}=200$. Left panel: the disconnected contribution to
            $\sigma_{\pi N}$ with a total of 56400 measurements;
            central panel: $\sigma_{s}$ with a total of 58560
            measurements; and right panel: $\sigma_{c}$ with a total
            of 58560 measurements. All results are obtained using the
            one-end trick.  \label{vsTSMPlots}}
      \label{vvTrickSigmaTSM}
    
  \end{figure*}

 \begin{figure*}[h!]
    
      \begin{minipage}{0.325\linewidth}
		\includegraphics[width=\linewidth,angle=0]{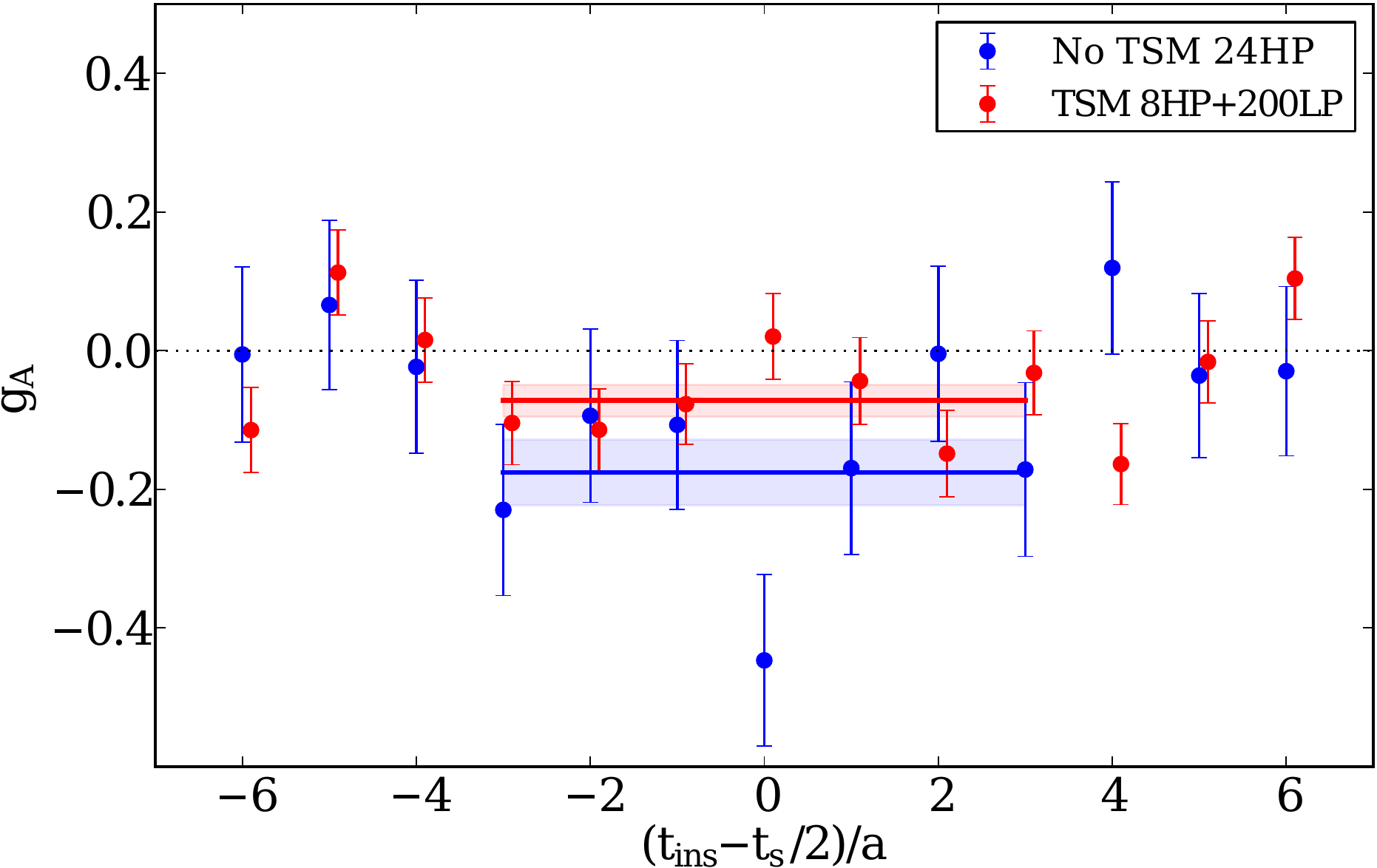}
      \end{minipage}
      \begin{minipage}{0.325\linewidth}
        \includegraphics[width=\linewidth,angle=0]{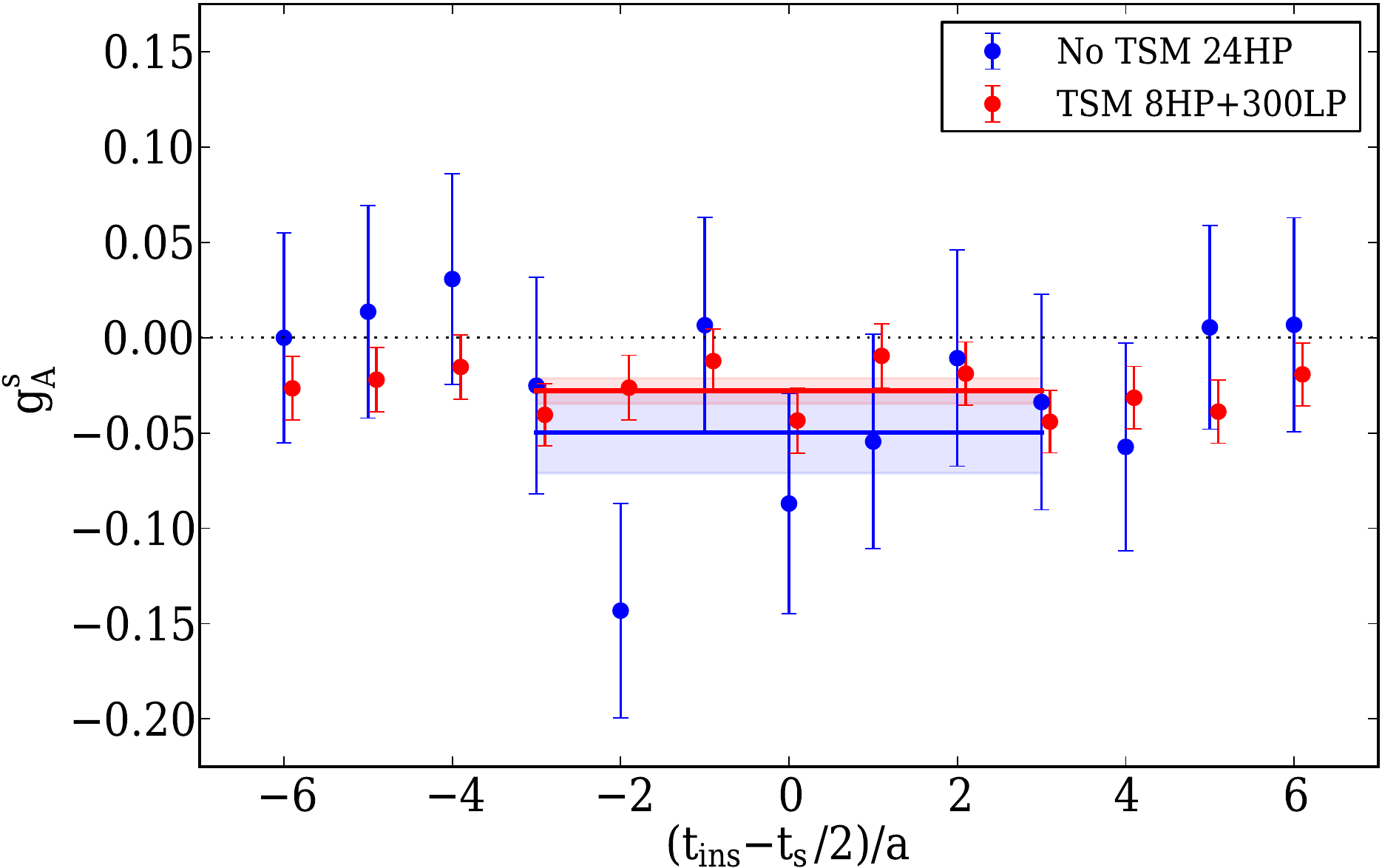} 
      \end{minipage}
      \begin{minipage}{0.325\linewidth}
        \includegraphics[width=\linewidth,angle=0]{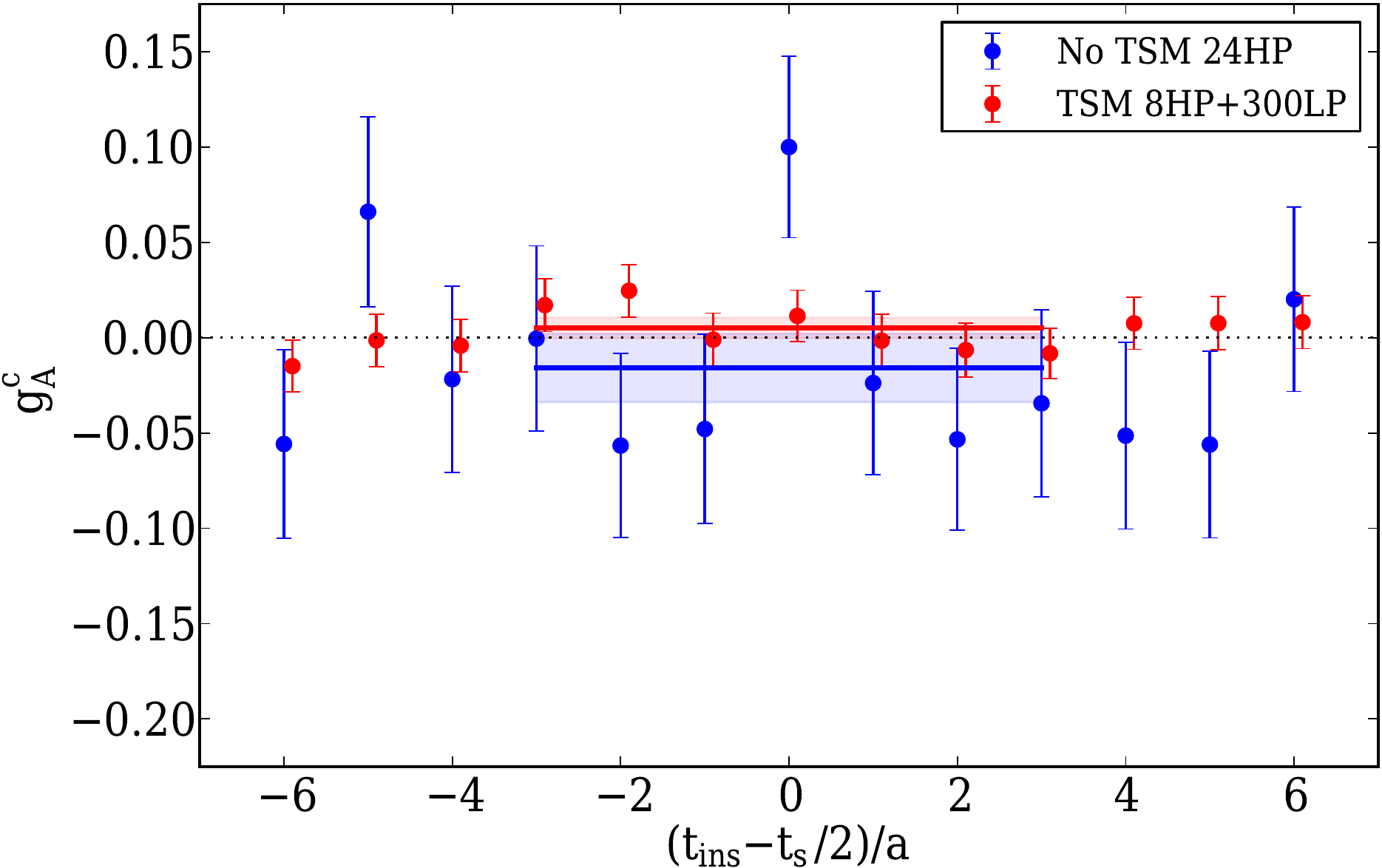} 
      \end{minipage}
	  \caption{The same as in Fig.~\ref{vvTrickSigmaTSM}
but for the disconnected contributions to the nucleon axial charge.}
	  \label{gAvsTSMPlots}
    
  \end{figure*}

  \begin{figure*}[h!]
    
      \begin{minipage}{0.4875\linewidth}
		\includegraphics[width=\linewidth,angle=0]{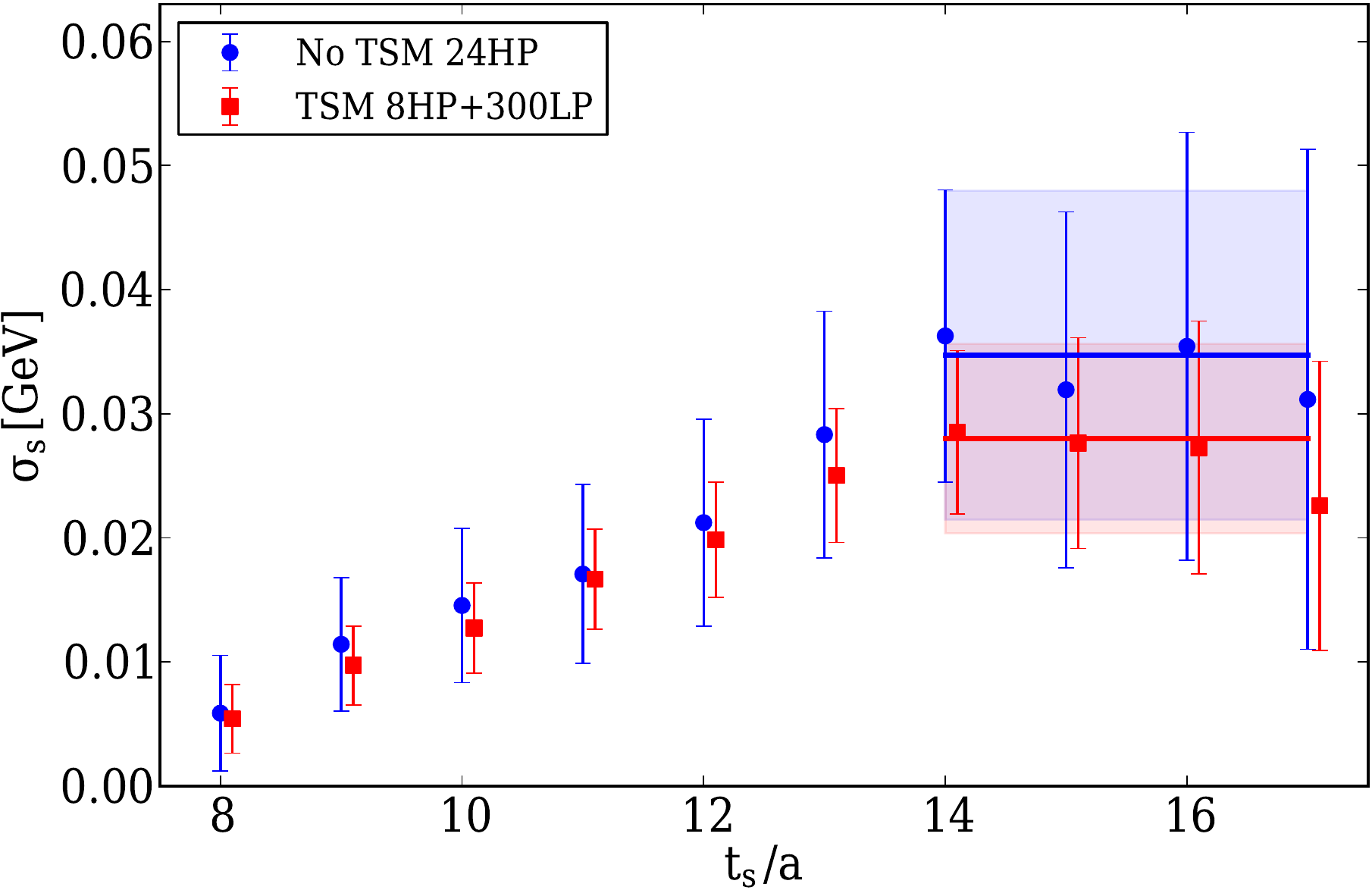}
      \end{minipage}
      \begin{minipage}{0.4875\linewidth}
        \includegraphics[width=\linewidth,angle=0]{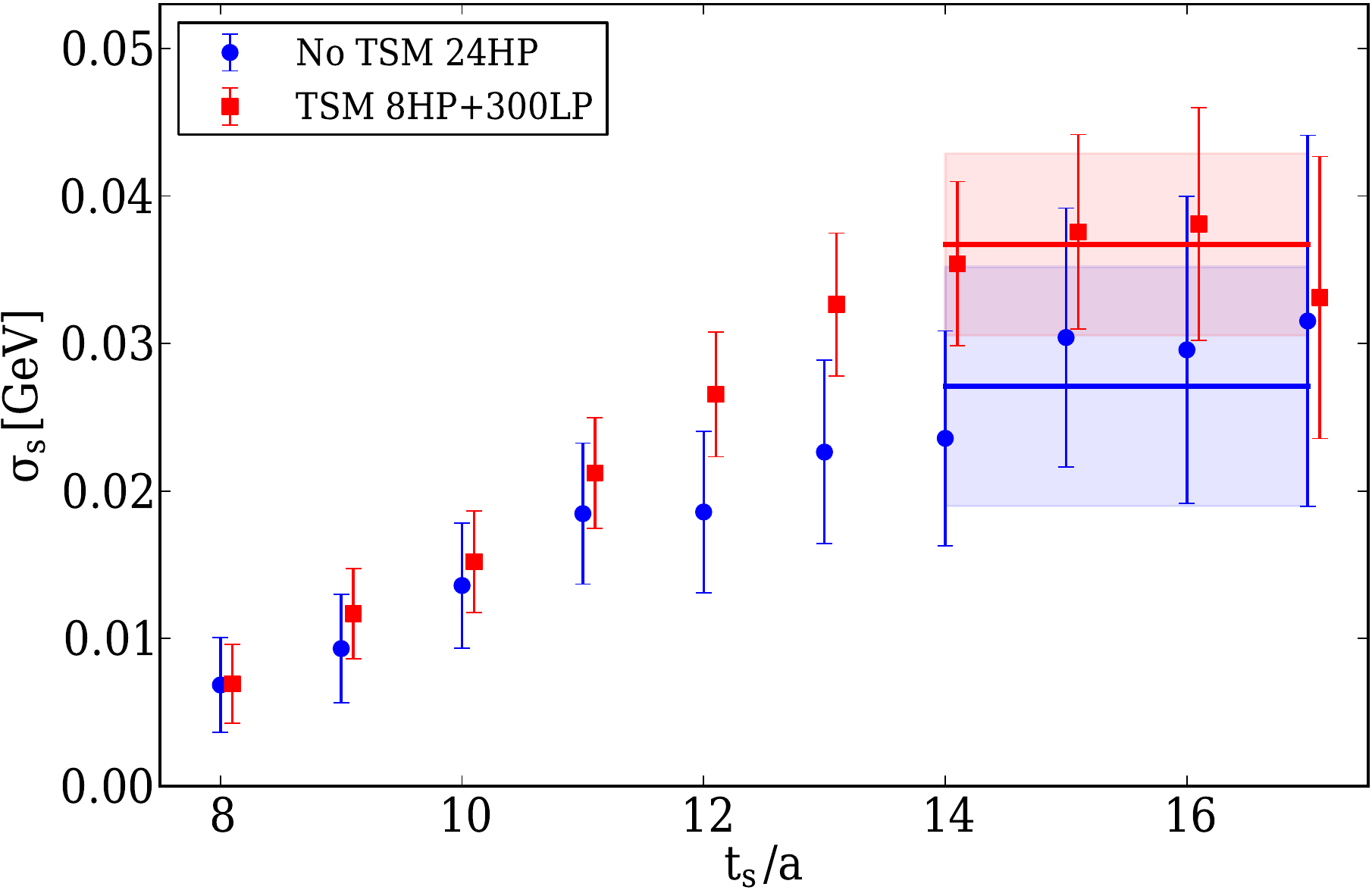} 
      \end{minipage}
	  \caption{Results for the ratio from which $\sigma_s$ is
            extracted versus the sink time separation $t_s$ when using
            only $N_{\rm HP}=24$ (no TSM) to those obtained when using
            $N_{\rm HP}=8$ and $N_{\rm LP}=300$.  Results are obtained
            using time-dilution (left panel) and time-dilution plus
            HPE (right panel).  In all cases the insertion-source
            separation $t_{\rm ins}= 8a$ and a total of 18628
            measurements are performed.  }
	  \label{tDilSigmavsTSMPlots}
    
  \end{figure*}

 \begin{figure*}[h!]
    
      \begin{minipage}{0.4875\linewidth}
		\includegraphics[width=\linewidth,angle=0]{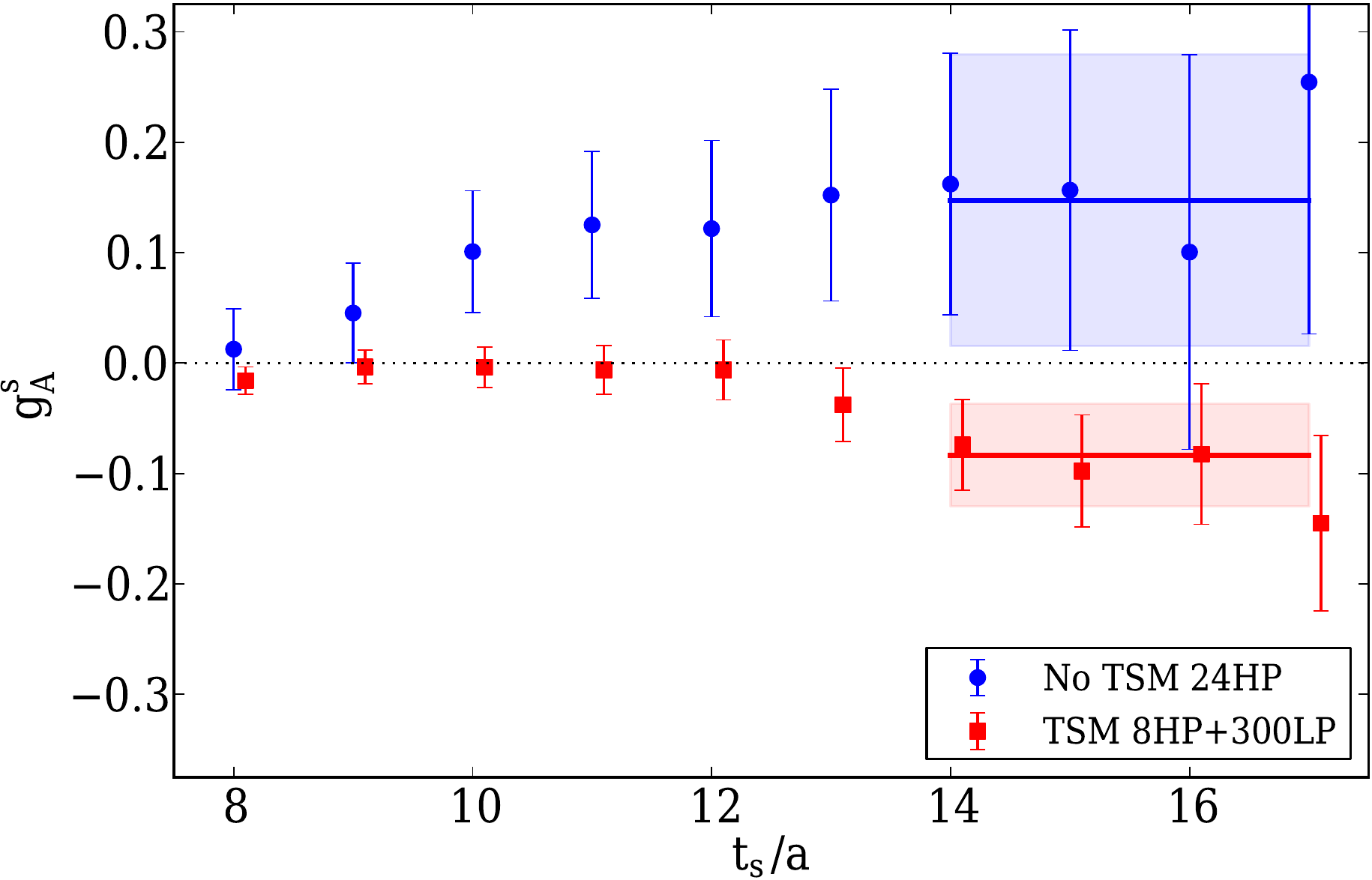}
      \end{minipage}
      \begin{minipage}{0.4875\linewidth}
        \includegraphics[width=\linewidth,angle=0]{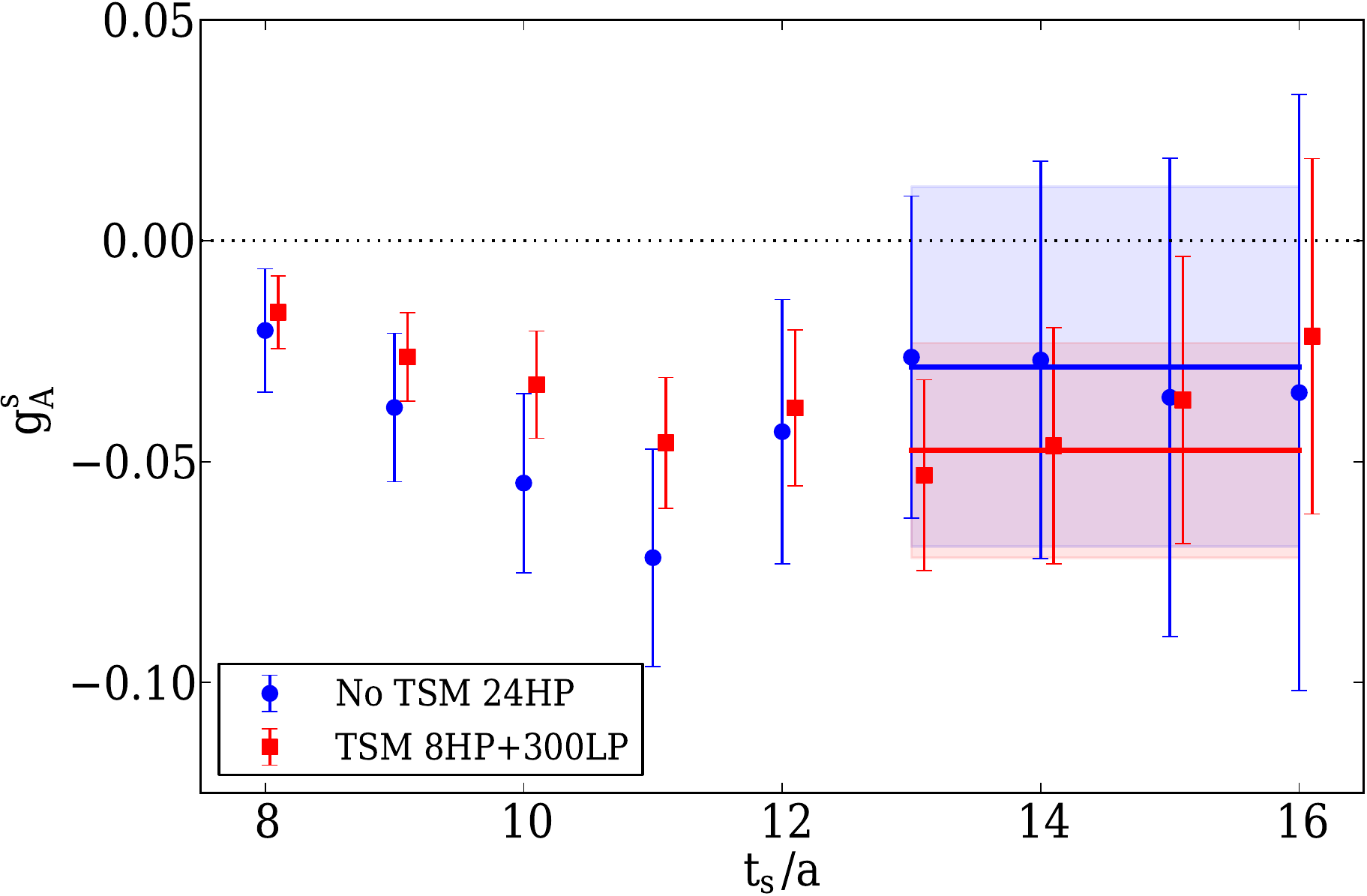} 
      \end{minipage}
	  \caption{The same as in Fig.~\ref{tDilSigmavsTSMPlots} but for the case of $g_A^s$.}
	  \label{tDilGAvsTSMPlots}
    
  \end{figure*}

  \begin{figure*}[h!]
    
      \begin{minipage}{0.4875\linewidth}
		\includegraphics[width=\linewidth,angle=0]{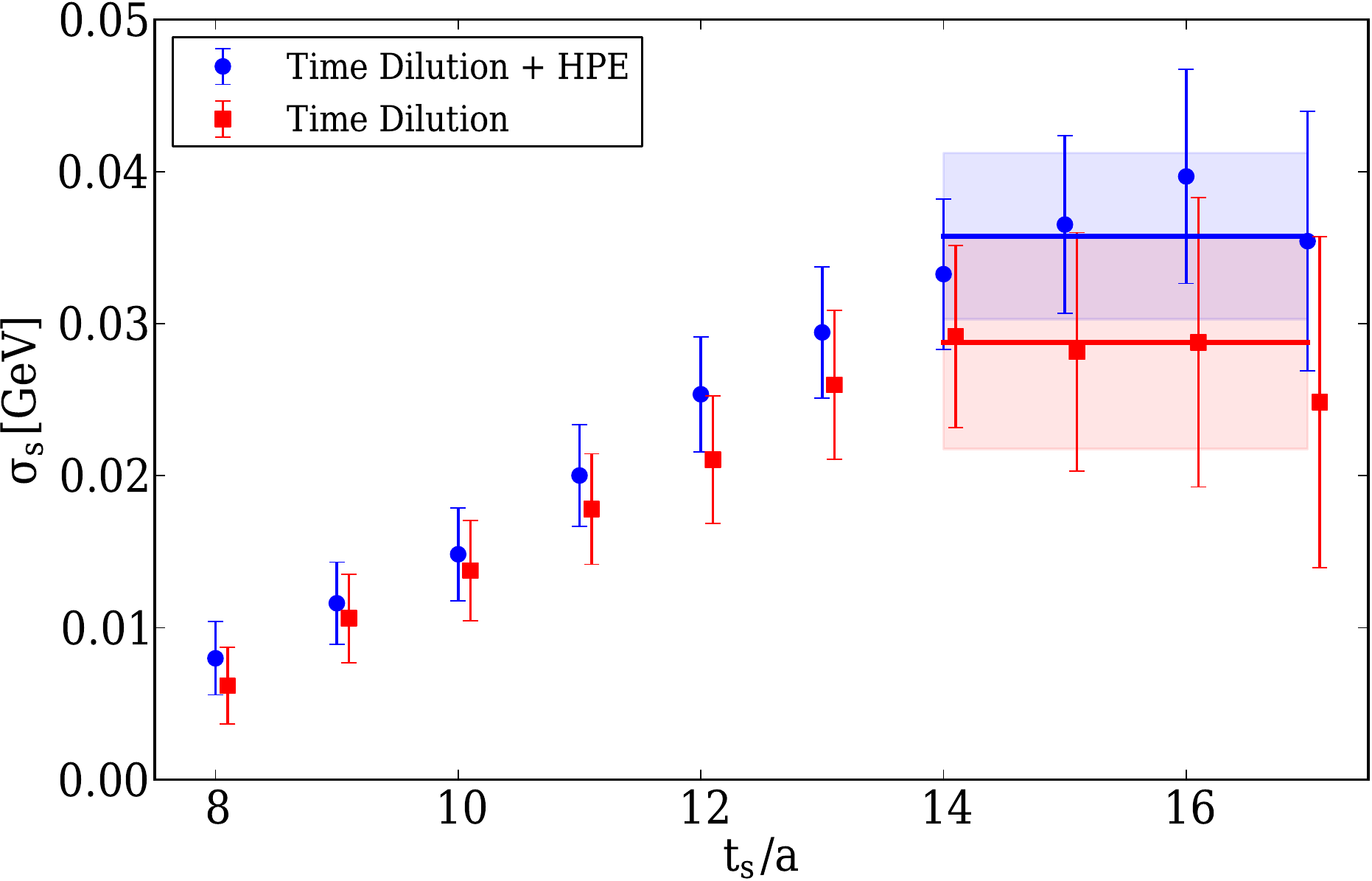}
      \end{minipage}
      \begin{minipage}{0.4875\linewidth}
        \includegraphics[width=\linewidth,angle=0]{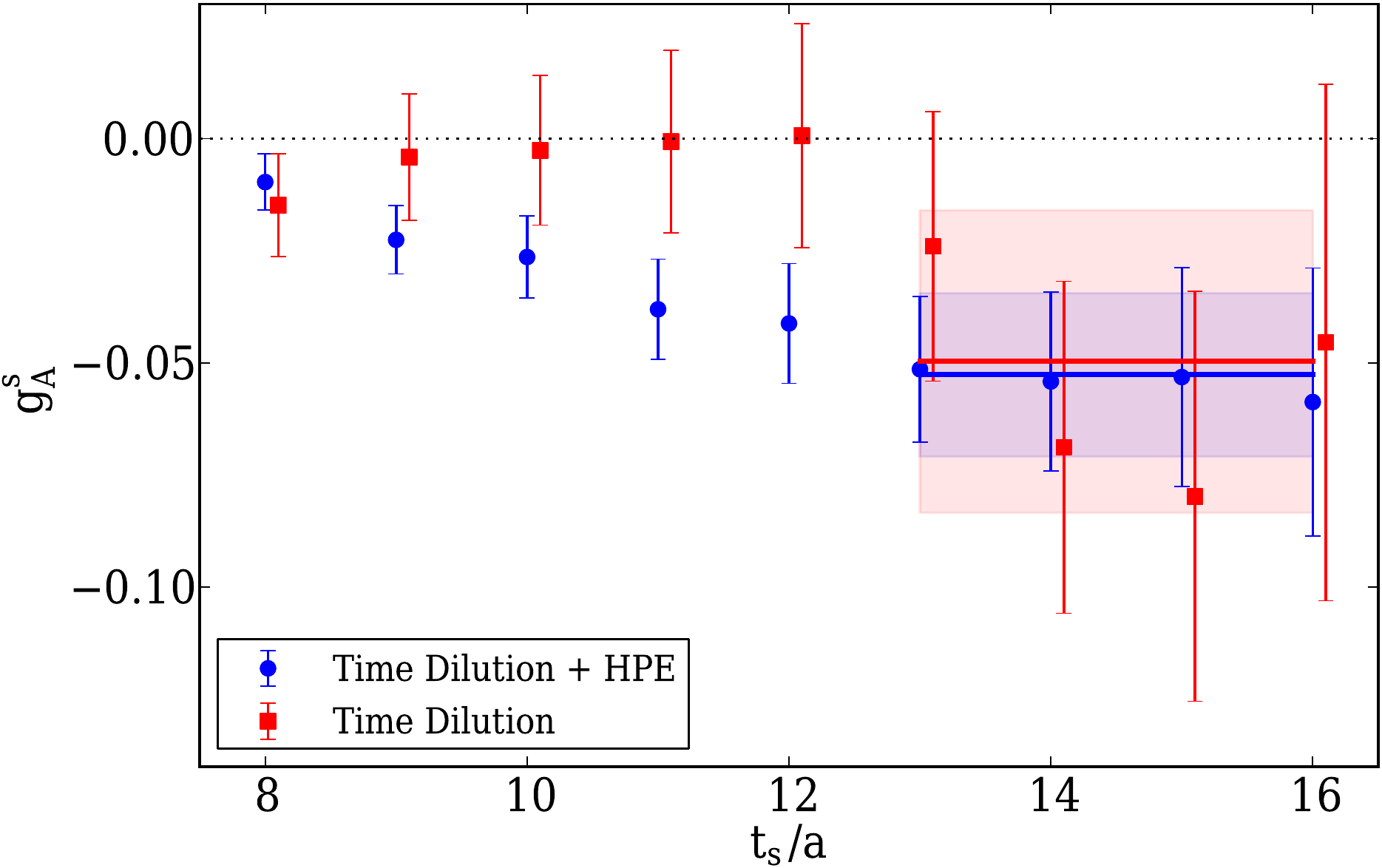} 
      \end{minipage}
	  \caption{Comparison of results for the ratio from which
            $\sigma_s$ (left panel) and $g_A^s$ (right panel) are
            extracted using time-dilution in combination with the TSM
            with and without application of the HPE. A total of 18628
            measurements are used. }
	  \label{tDilVsHPE}
    
  \end{figure*}

 \begin{figure*}[h!]
    
      \begin{minipage}{0.4875\linewidth}
		\includegraphics[width=\linewidth,angle=0]{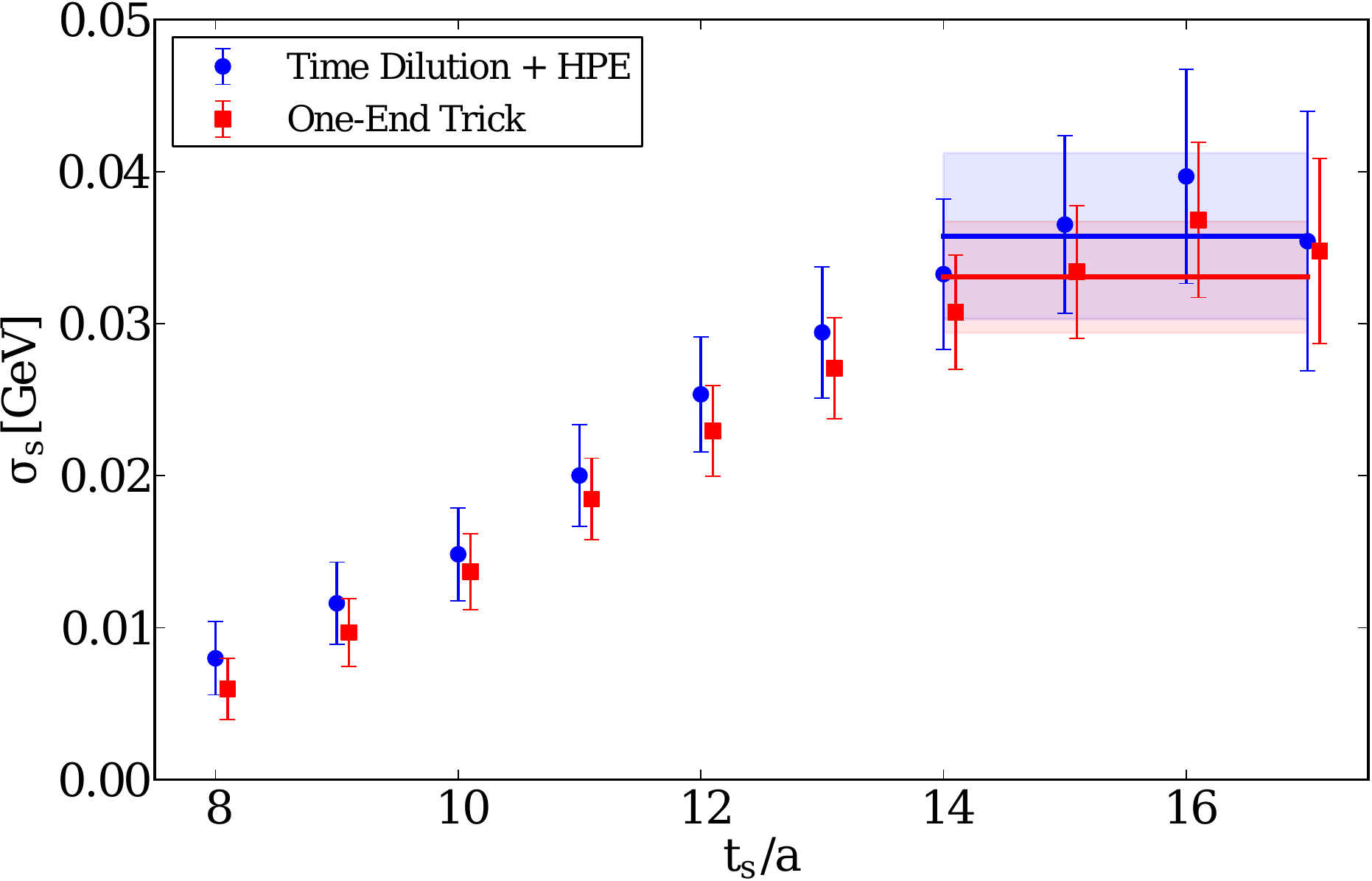}
      \end{minipage}
      \begin{minipage}{0.4875\linewidth}
        \includegraphics[width=\linewidth,angle=0]{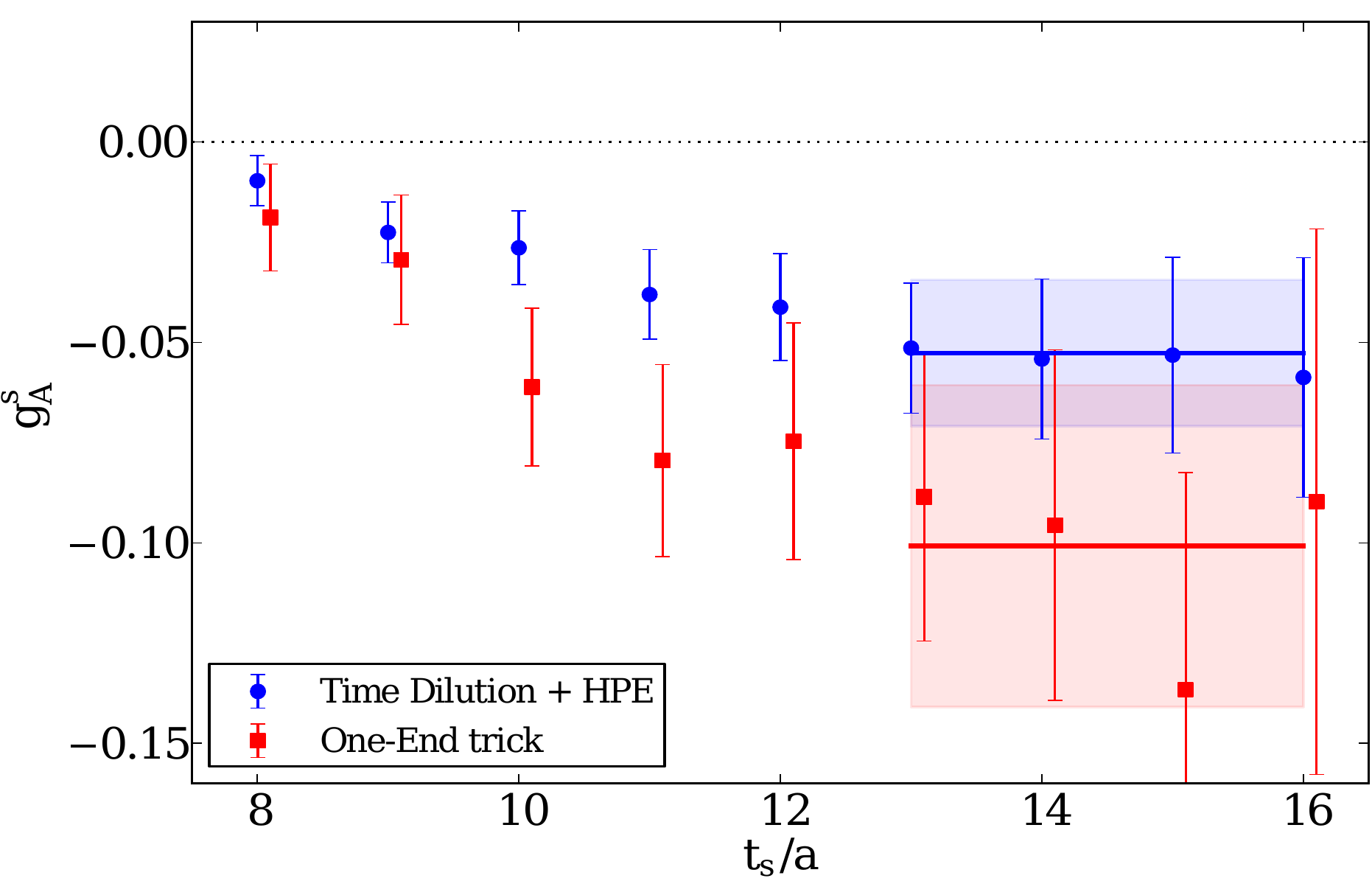} 
      \end{minipage}
	  \caption{Results for the ratio from which $\sigma_s$ (left)
            and $g_A^s$ (right) are extracted. With filled (blue)
            circles are results obtained using the one-end trick and
            with filled (red) squares  when using time-dilution. In
            both cases we use the TSM with $N_{\rm HP}=24$ and $N_{\rm
              LP}=300$ and 18628 measurements. The current insertion
            is fixed at $t_{\rm ins}=8a$.}
	  \label{tDilVsVV}
    
  \end{figure*}
 \begin{figure}[h!]

    		\includegraphics[width=\linewidth,angle=0]{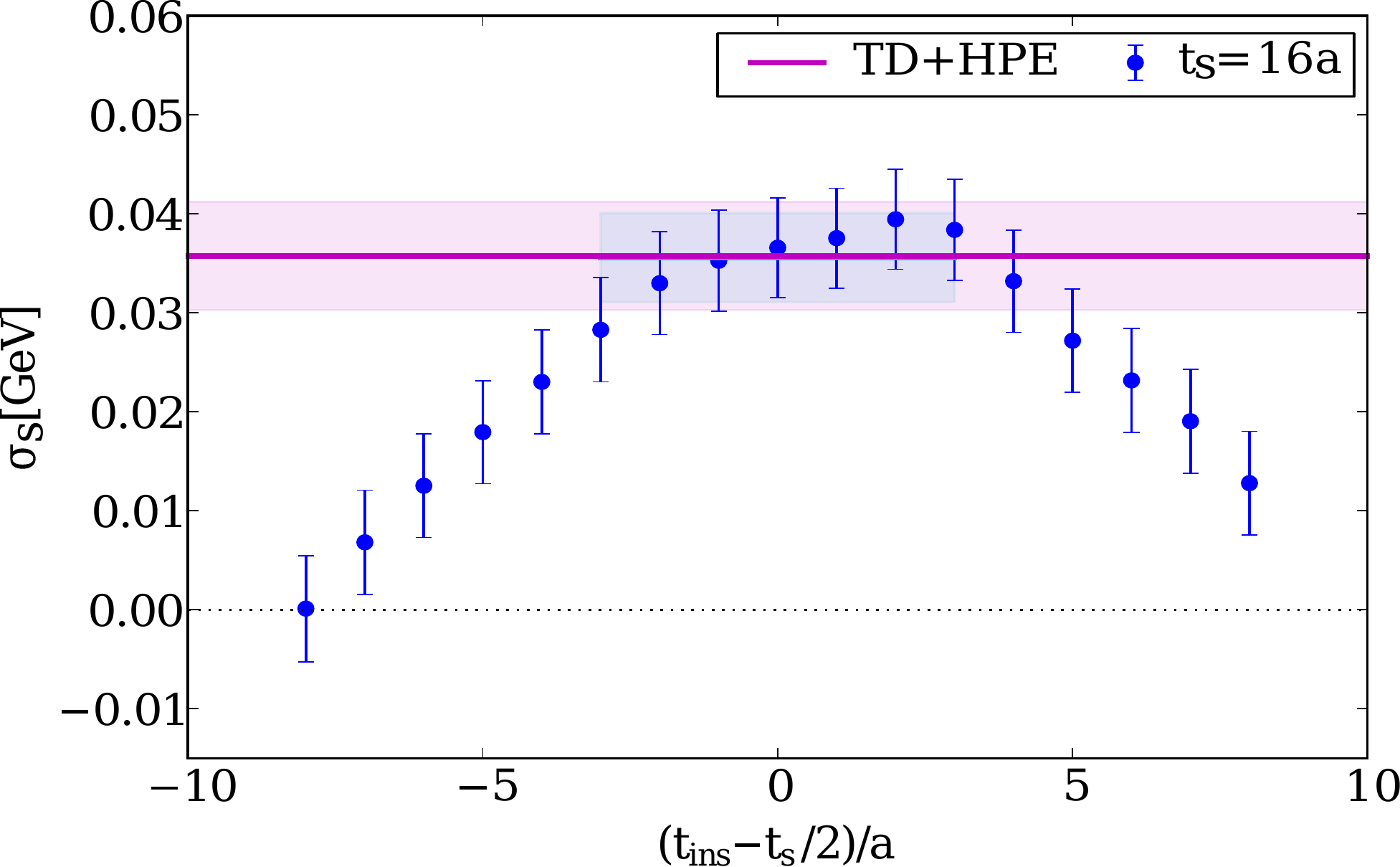}\\
        \includegraphics[width=\linewidth,angle=0]{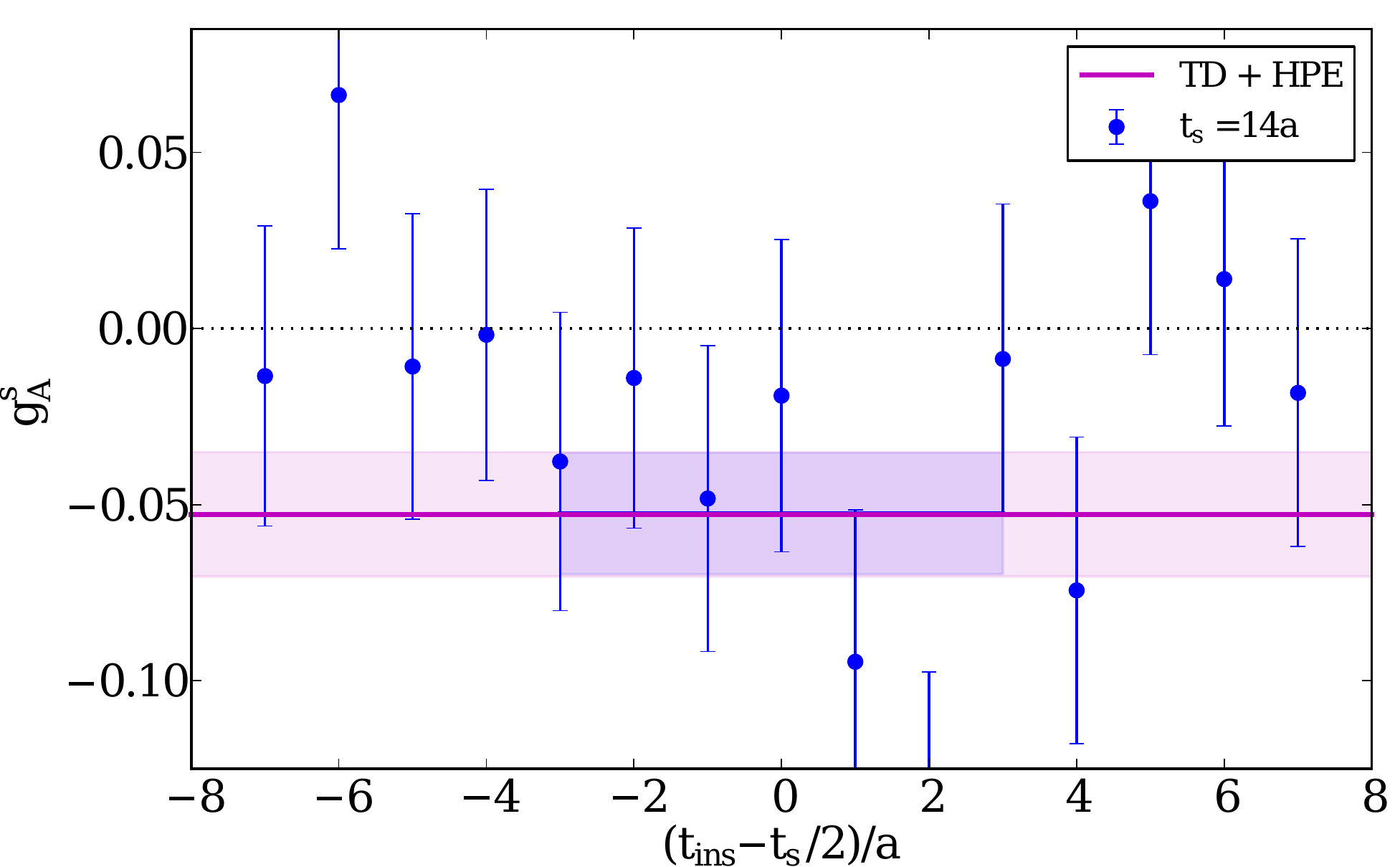} 
	  \caption{Ratios for $\sigma_s$ at $t_s=16a$ (top) and
            $g_A^s$ at $t_s=14a$ (bottom) obtained using the one-end
            trick. The purple band shows the result of fitting the
            asymptotic behavior of the ratio obtained with
            time-dilution. The TSM with $N_{\rm HP}=24$ and $N_{\rm
              LP}=300$ is used in both methods with 18628 statistics.}
	  \label{PlotVsVV2}
    
  \end{figure}

We first examine the performance of TSM for the $\sigma$-term.  In
Fig.~\ref{vvTrickSigmaTSM} we show the disconnected contributions for
$\sigma_{\pi N}$, the strange $\sigma_s=\mu_s\langle N |\bar{s}s
|N\rangle$ and charm $\sigma_c=\mu_c\langle N|\bar{c}c|N\rangle$
nucleon $\sigma$-terms. The strange and charm $\sigma$-terms are
computed using Osterwalder-Seiler fermions with $\mu_s$ and $\mu_c$
tuned to reproduce the kaon and D-meson masses of the unitary
theory. Results are obtained using the one-end-trick with and without
applying the TSM. For the case where we employ TSM, we use $N_{\rm
  LP}=200$ for loops containing light quarks and $N_{\rm LP}=300$ for
strange and charm quark loops. These choices for $N_{\rm LP}$ yield
approximately the same statistical errors allowing a more direct
comparison of computer time. Namely, for the case of $\sigma_{\pi N}$,
we obtain results with similar errors but with reduced computational
cost for the TSM by $\sim 34\%$ showing that the TSM is preferable. As
the quark mass increases, the computational cost for the TSM for
similar errors becomes comparable to that of using only HP 
inversions. Thus for $\sigma_s$ the TSM is comparable to only using
$N_{\rm HP}=24$. For even heavier masses such as in the case of the
charm quark the use of the TSM is not justified since the computer
time increases by a factor of 5, while the errors are reduced by a
mere $\sim 33\%$. Thus, when the inversion of the Dirac matrix is fast
as in the case of the charm quarks there is not much benefit from
using lower precision. Rather the increased number of contractions
required when using the TSM, which is a constant overhead independent
of the quark mass, becomes more significant than any speed-up obtained
by using lower-precision inversions.

We perform the same analysis for $g_A$, which has a different
convergence pattern as compared to the $\sigma$-terms.  Contrary to
the case of the $\sigma$-terms, for $g_A$ one must use the generalized
version of the one-end trick since computing the isoscalar axial
charge in the twisted basis requires summing the quark-flavor
contributions.  In Fig.~\ref{gAvsTSMPlots} we show results for the
disconnected light quark contributions to $g_A$, the strange and charm
contributions to the nucleon axial charge denoted by $g_A^s$ and
$g_A^c$ respectively. As can be seen, there is an improvement when
using the TSM for all quark masses, though the improvement is more
significant the lighter the quark mass is. In the most favorable case,
i.e. that of the light quark sector, we see more than a two-fold
reduction in the error when using the TSM for about 66\% of the
computational cost.  In the case of $g_A^c$, although the TSM is
computationally more demanding by a factor of $5$ for the chosen
parameters of the plot, the four-fold reduction in the error
over-compensates for this cost.

We next assess the performance of the TSM in combination with
time-dilution instead of with the one-end trick as done above for the
same two observables considered.  The comparison is performed for the
strange quark fermion loops in order to speed-up the computations.
Time-dilution also allows straightforward application of the HPE
method, which potentially can lead to improvement in particular for
heavier quark masses.  As already explained, the overhead in computer
time when applying the HPE is insignificant, since it essentially
requires a few applications of the Wilson-Dirac operator.  In
Fig.~\ref{tDilSigmavsTSMPlots} we show the results for the ratio from
which $\sigma_s$ is extracted using $N_{\rm HP}=24$ high precision
inversions and $N_{\rm LP}=0$ compared to those obtained when using
the TSM with $N_{\rm HP}=8$ and $N_{\rm LP}=300$.  The computational
cost in the two cases is roughly the same.  As can be seen, the TSM
yields smaller errors by about a factor of two both with and without
the HPE.  For the case of $g_A^s$ shown in Fig.~\ref{tDilGAvsTSMPlots}
the results are even more favorable for the TSM, where one 
obtains the right long time behavior even when the HPE is not applied.
 Using time dilution with $N_{\rm HP}=24$ only we obtain the wrong results indicating that no convergence has been reached.
  The TSM yields better than a two-fold reduction in
errors for the same computer time yielding results consistent with
those obtained using the one-end trick. Thus, applying the HPE leads
to improvement and it should be employed when using time-dilution.

It is helpful to directly compare the results obtained with
time-dilution and the TSM with and without the HPE.  As explained,
applying the HPE comes with almost no computational cost.  A direct
comparison is shown in Fig.~\ref{tDilVsHPE}. As can be seen, errors
are reduced by about a factor of two in the case of $g_A^s$ when using
the HPE.  Moreover, we expect a greater improvement as the quark mass
becomes heavier. Since the addition of HPE improves results without
increasing the computer time in a noticeable way, it is always
advantageous to use it for quark masses in the range of the strange
quark or heavier.

It is important to stress that the creation of stochastic sources,
the inversions and all contractions are carried out 
on GPUs such that the communication between CPU and GPU is reduced. In order
to do that, the sources are directly contracted on the GPUs right after
the inversion, the calculated propagators are discarded, and only the
contractions are transferred to the CPUs to be stored on disk~\cite{alex}.

Even with such a setup, for quark masses larger than the strange
quark mass the differences in computer time between high and low precision inversions
  become small 
as compared to the time spent for contractions to calculate the loop.
 This is due to the fact that the pre- and post-processing
computational costs are independent of the quark mass and therefore
more time consuming for the TSM where an order of magnitude more
noise vectors are used, thus reducing the improvements observed by
the TSM for the case of heavy quarks.
In Table~\ref{compRatios} we give a summary of the computer time
 required for the computation of fermion loops within the various methods. 
We give the ratio $R_{\rm HP/LP}$ of the computer time required to
 compute a  fermion loop for one noise vector
using HP to the  time needed to compute the loop
using  LP,  taking into account the time for the inversion as well
as the pre- and post-processing time (creation of sources, performing the
  contractions and taking the traces). A large value for this ratio indicates that the TSM is
more efficient, since more LP vectors can be used for the
computation of the  loops as compared to  the cost of a loop using one
HP inversion. A value close to unity
indicates that the TSM is no longer advantageous, since in such a case one can
exchange LP inversions for HP with the same cost.
 For the case of the 
strange quark loops with
local operator insertion $R_{\rm HP/LP}\gg 1$ when  time-dilution is applied
 either with or without HPE. In the case of using the one-end trick to compute
the fermion loops the TSM 
has a better performance for both ultra-local and one-derivative operator insertions for light and strange quarks.
As the quark  mass increases, $R_{\rm HP/LP}$ decreases making 
 the TSM less advantageous for charm quarks.

\begin{table}[h!]

\begin{tabular}{|c|c|c|c|c|c|}
\hline
   Method       & Quark sector & $R_{\rm HP/LP}^{Local}$ & $R_{\rm HP/LP}^{One-Deriv.}$ \\
\hline
One-end trick   &    Light     &       $\sim26.7$        & $\sim10$ \\
One-end trick   &    Strange   &       $\sim16.9$        & $\sim5.8$ \\
One-end trick   &    Charm     &       $\sim 2.9$        & $\sim1.4$ \\
Time-dil.       &    Strange   &       $\sim20.7$        & --- \\
Time-dil. + HPE &    Strange   &       $\sim19.1$        & --- \\
\hline 
\end{tabular}
\caption{Computational cost when using TSM with the one-end trick or
  with time-dilution for different quark masses. The third column is
  the ratio of the cost for computing a fermion loop using a HP
  inversion to a low precision one, including inversion time and time
  for pre- and post-processing for ultra-local operator
  insertions. The fourth column gives the corresponding ratio when
  including one-derivative operators to the ultra-local
  ones.}\label{compRatios}

\end{table}

We note here that we  have not carried out an analysis of time-dilution for the case of
derivative insertion operators, since  one would require
to include fermion loops computed at three time-slices to take the time derivatives,
which would effectively triple the cost of time-dilution. 

Our main conclusion from the comparison carried out in this section is that the 
 TSM is the method of choice for 
light quarks and  for the case of operators where  the generalized one-end
trick is used. In the charm quark mass range 
with our current implementation on GPUs the TSM becomes
less efficient since  the pre- and post-processing overheads become large
as compared  to the inversion time. For observables  like $g_A$ the TSM is still
superior  for computing strange quark loops and remains
  an equally good option for charm quark loops. For the $\sigma$-terms the
one-end trick works very well and the TSM is not necessary. However,
since our goal is to  compute all loops at once   the TSM is the method of choice for 
obtaining high statistics results if one wants to 
compute all  the disconnected contributions
to observables probing nucleon structure.

%
%


\subsection{Time-dilution plus HPE vs the one-end trick}

In the previous section we compared results obtained using the one-end
trick as well as time-dilution with and without the TSM and the HPE.
Here we employ the TSM with $N_{\rm HP}=24$ and $N_{\rm LP}=300$ and
compare results obtained with the one-end trick to those obtained
using time-dilution with HPE.  In Fig.~\ref{tDilVsVV} we show results
for the ratio from which $\sigma_s$ and $g_A^s$ are extracted. The
ratio is plotted as a function of the sink-source separation $t_s$
for fixed current insertion time $t_{\rm ins}=8a$.  In the case of
$\sigma_s$ results obtained using the one-end trick of
Eq.~\eqref{loopVv} are compared to those obtained using time-dilution
and HPE, whereas for $g_A^s$ the generalized one-end trick of
Eq.~\eqref{loopStD} is compared to time-dilution and HPE.  As can be
seen, for $\sigma_s$ the one-end trick yields smaller errors  than
time-dilution for the same statistics.  On the other hand, for $g_A^s$
time-dilution yields smaller errors.  However, in the case of the
one-end trick one obtains the fermion loops at all time-slices without
any further inversions, while when using time-dilution the fermion
loop is calculated at a single time-slice or at up to four in our
setup when using the coherent source method.  As a consequence, with
the one-end trick we can obtain results for all current insertions
{\it and} for multiple sink-source time separations.  Thus one can fit
the plateau as shown in Fig.~\ref{PlotVsVV2} and compare with the
result extracted when using time-dilution at fixed $t_{\rm ins}$.
Fitting the plateau for $g_A^s$ yields a result with the same error as
that obtained using time-dilution.  Thus, this comparison shows that
the one-end trick is preferable for the calculation of fermion loops
even when the generalized form of Eq.~\eqref{loopStD} is used.

\begin{figure}[h!]
  
    \includegraphics[width=0.87\linewidth,angle=0]{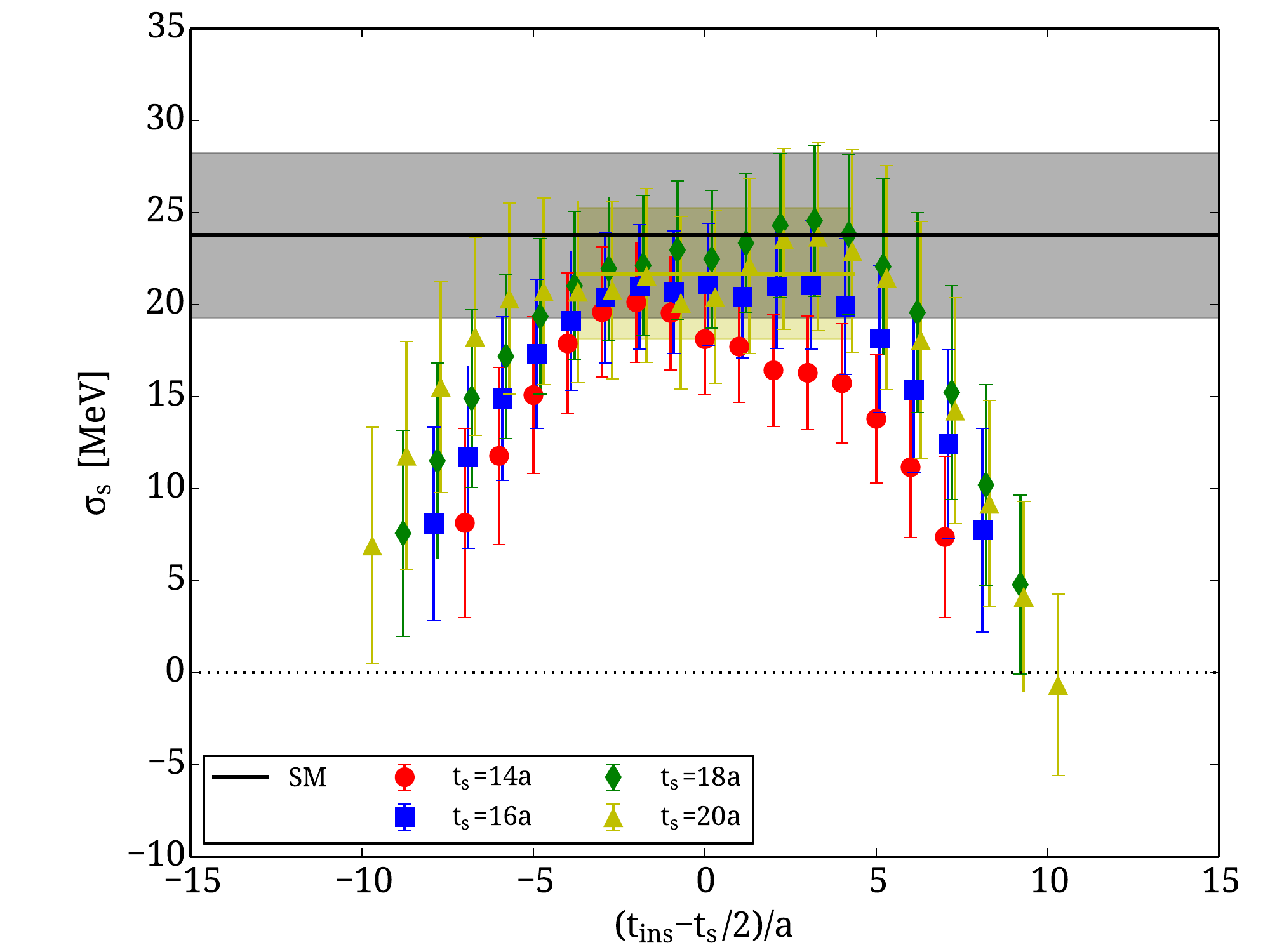}\\
    \includegraphics[width=0.87\linewidth,angle=0]{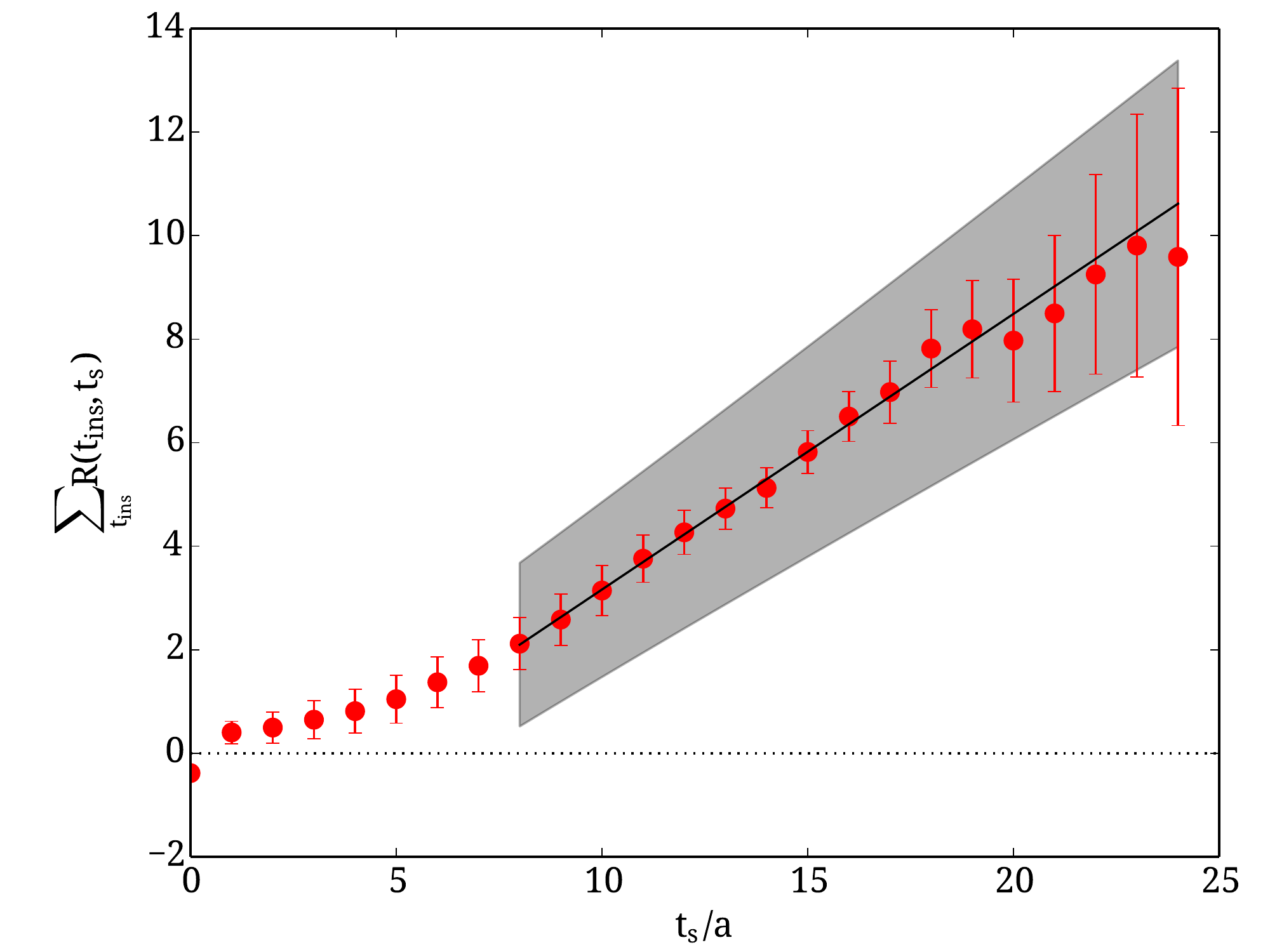}\\
    \includegraphics[width=0.87\linewidth,angle=0]{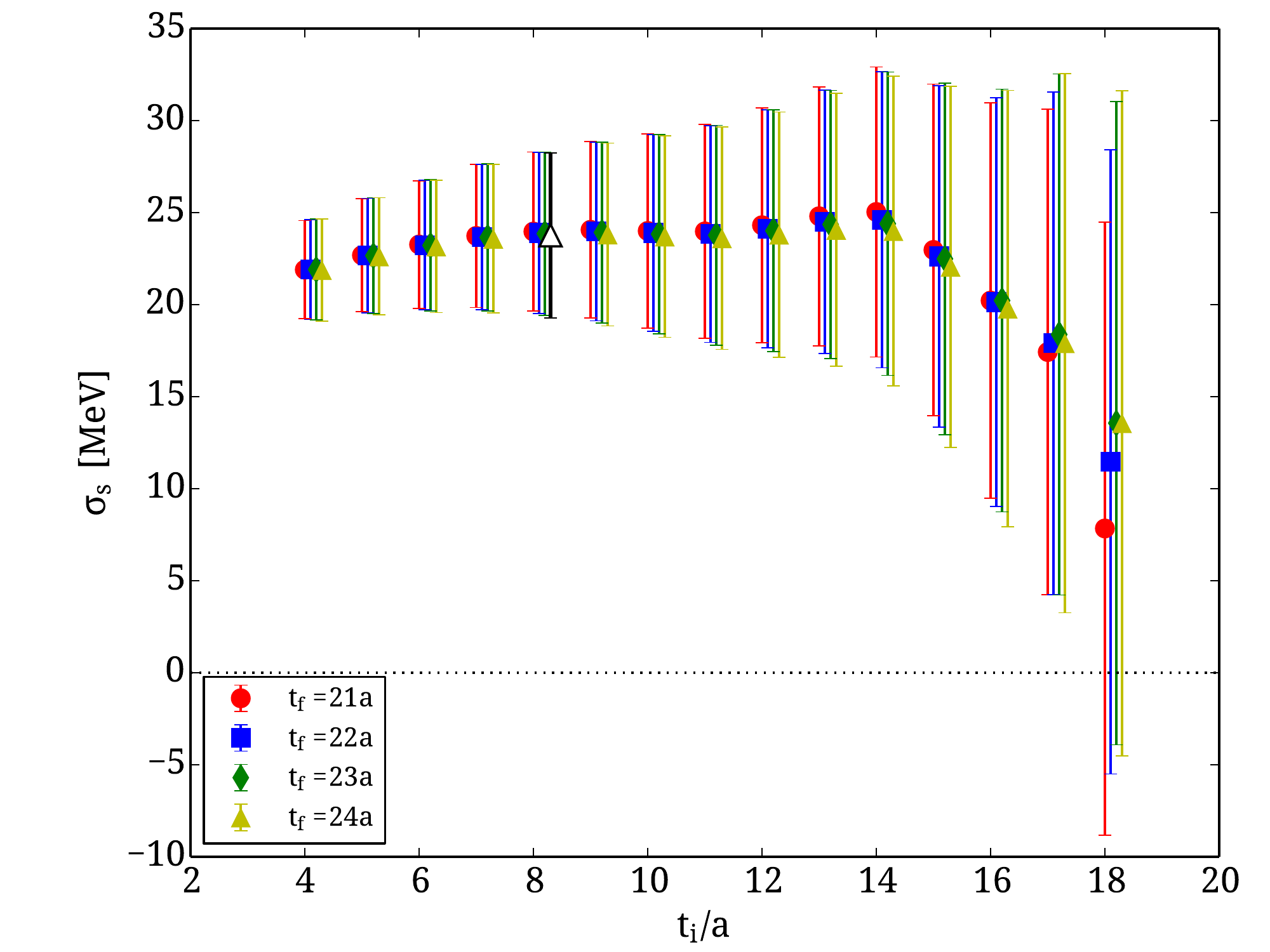} 
      \caption{Comparison of the summation and the plateau methods for
         $\sigma_s$. In the upper panel we show the
        ratio as a function of the insertion time-slice with respect
        to mid-time separation ($t_{\rm ins}-t_s/2$) for source-sink
        separations, $t_{\rm s}=$14$a$ (red circles), $t_{\rm s}=16a$
        (blue squares), $t_{\rm s}=18a$ (green rhombuses) and $t_{\rm
          s}=20a$ (yellow triangles).  In the middle panel we show the
        summed ratio, for which the fitted slope yields the desired
        matrix element. In the bottom panel we show the results
        obtained for the fitted slope of the summation method for
        various choices of the initial and final fit time-slices. The
        open triangle shows the choice for which the gray bands are
        plotted in the upper and middle panels.
        \label{sigmaStrange}}
  
\end{figure}

\begin{figure}[h!]
  
    \includegraphics[width=0.87\linewidth,angle=0]{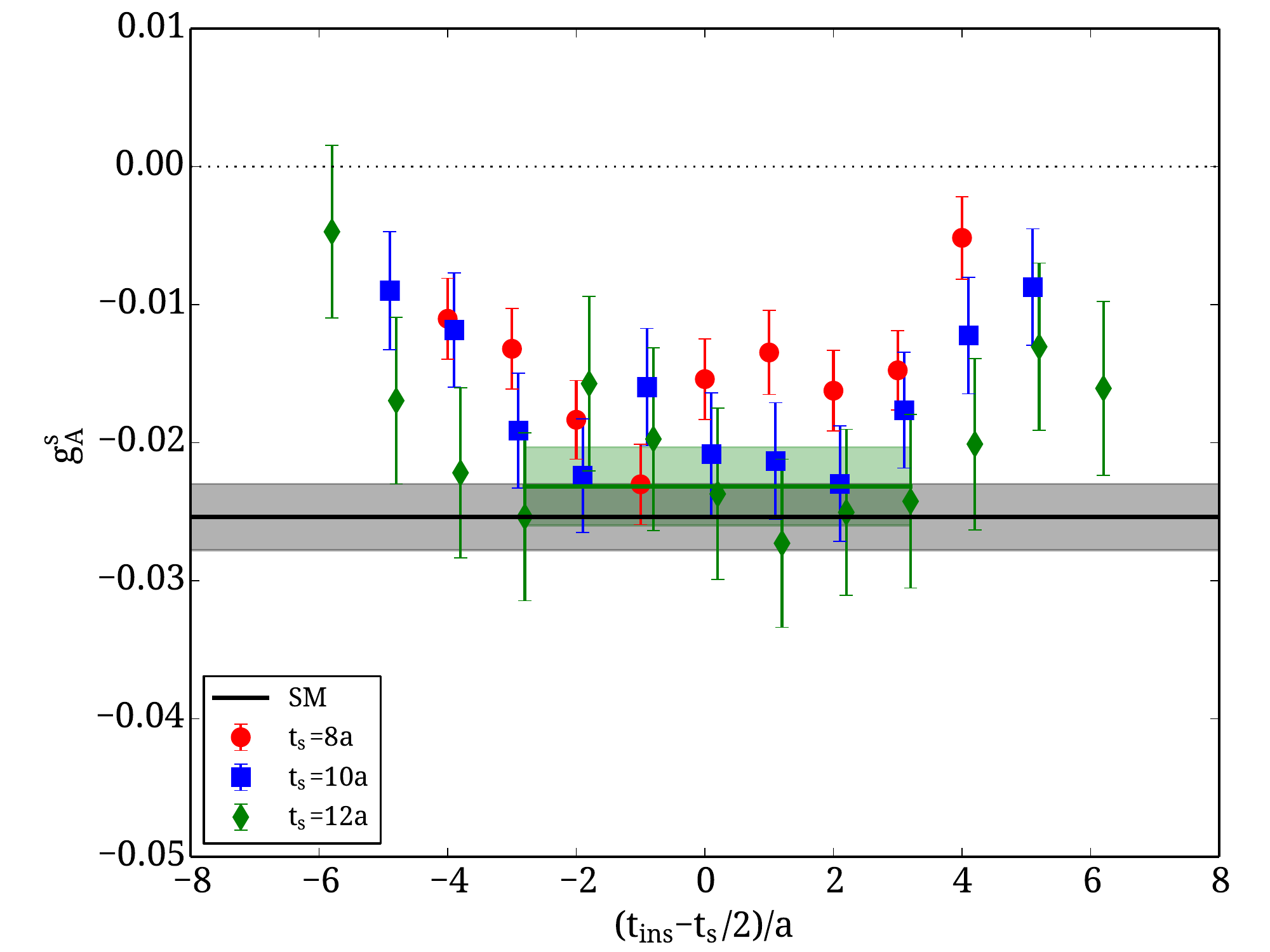} \\
    \includegraphics[width=0.87\linewidth,angle=0]{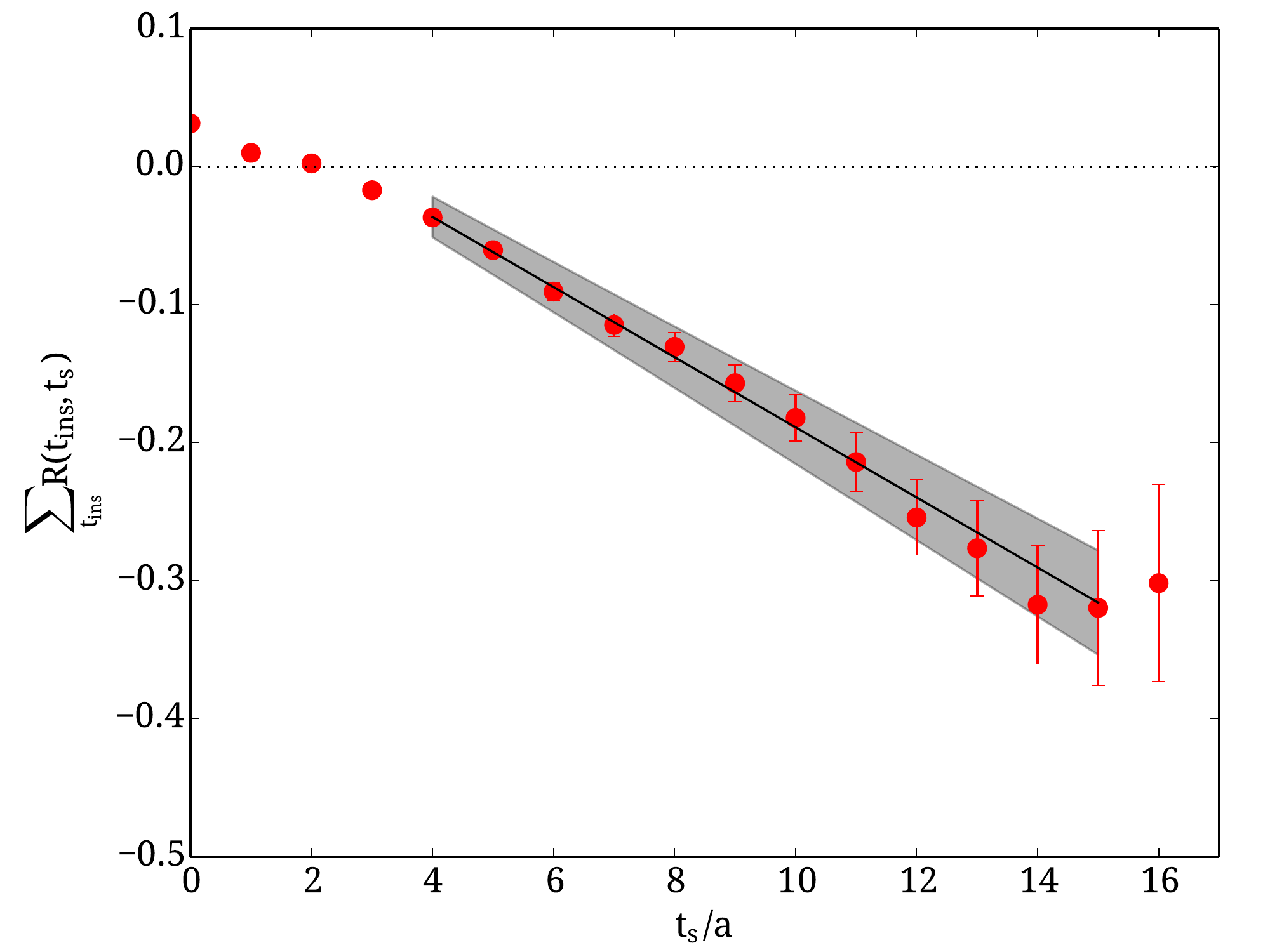} \\
    \includegraphics[width=0.87\linewidth,angle=0]{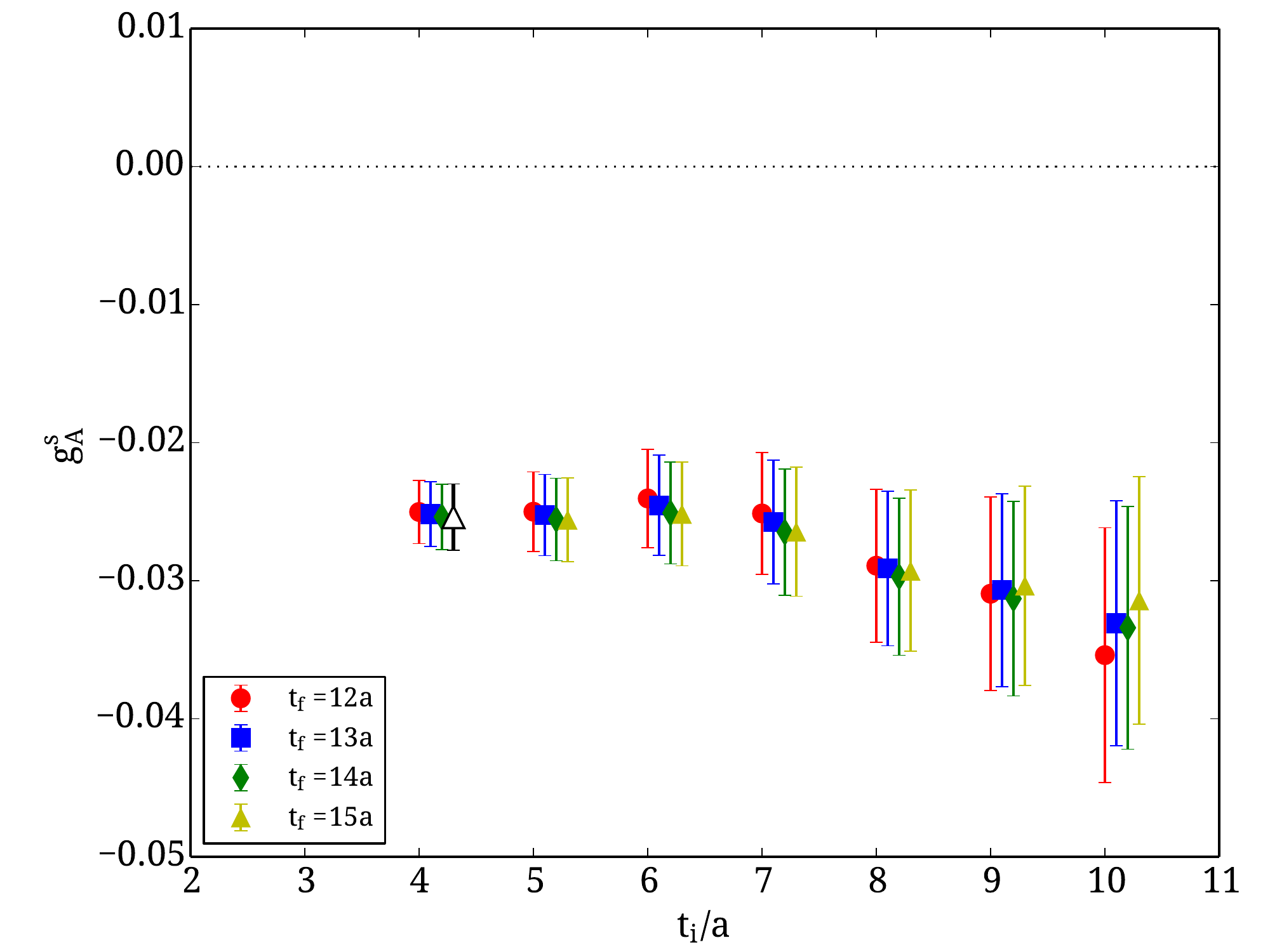}
    \caption{Comparison of the summation and the plateau methods for
      the strange contribution to the axial charge: $g_A^s$. In the
      upper panel we show the ratio as a function of the insertion
      time-slice with respect to mid-time separation ($t_{\rm
        ins}-t_s/2$) for source-sink separations, $t_{\rm s}=$8$a$
      (red circles), $t_{\rm s}=10a$ (blue squares) and $t_{\rm
        s}=12a$ (green rhombuses). The rest of the notation is the
      same as that of Fig.~\ref{sigmaStrange}.\label{gAStrange}}
    
\end{figure}
 An additional advantage of the one-end trick is that having results
 for multiple sink-source time separations allows for the assessment
 of excited state contributions as well as for applying the summation
 method with no extra inversions.  In contrast, time-dilution requires
 an inversion for every new insertion time, which would effectively
 multiply the computational cost by the number of time-slices between
 source and sink of the largest separation considered. Furthermore,
 with the one-end trick one has the loop at all time-slices which
 allows coupling the loop to multiple two-point functions computed
 with different source positions. The two-point functions at each new
 source position require new inversions, however the loops are
 computed once with the one-end trick at all time-slices, thus
 multiplying the number of statistics at the cost of regular
 point-to-all inversions. The advantage of having multiple
 sink-sources separations is demonstrated for the strange
 $\sigma$-term ($\sigma_s$) and strange-quark contribution to the
 axial charge ($g_A^s$) shown in Figs.~\ref{sigmaStrange}
 and~\ref{gAStrange} respectively. In both cases we computed 16
 two-point functions per configuration on 2,300 gauge-field
 configurations resulting in 147,200 statistics since we average
 forwards and backwards propagating nucleons and proton and neutron
 channels. For this high statistics analysis we take $N_{\rm HP}=24$
 and $N_{\rm LP}=300$. 
As can be seen
 in Figs.~\ref{sigmaStrange} and \ref{gAStrange}, the multiple
 sink-source time separations are crucial in probing excited states
 contamination.  The summation method, which serves as a different way
 of extracting the observable, can only be applied if we have these
 multiple sink-source time separations. Although a noticeable
 improvement in statistical accuracy is not obtained when using the
 summation method, it is very useful as an additional check of
 convergence to the ground state, especially for the case of the
 $\sigma$-term where excited state effects appear to be larger.

We have carried out a comparison between time-dilution with HPE and
the one-end trick only for strange quark loops. We expect the one-end
trick to perform better for light quarks since HPE is less suited,
while for heavier masses time-dilution combined with HPE may become
advantageous due to the HPE. Another reason to favor the one-end trick
method is for the case of one-derivative operators. To compute such
derivative operators in time one requires the fermion loops at at
least three neighboring time-slices. For the one-end trick this
requires no further inversions since one obtains the loops at all
time-slices, however for time-dilution, where an inversion is required
at every time-slice, this triples the computational cost.

 \begin{table}[h!]
\footnotesize

\begin{tabular}{|c|c|c|c|c|}
\hline
	Method			&	Abs. Error	&	OH	&	Cost 	&	Cost$\times$Error$^2$	\\
\hline
\multicolumn{5}{|c|}{$\sigma_{\pi N}$ }				\\
\hline
One-end trick			&	4.3 MeV		&	65	&	2234	&		0.032		\\
One-end trick + TSM		&	3.8 MeV		&	290	&	1471	&		0.027		\\
\hline
\multicolumn{5}{|c|}{$\sigma_{s}$ }					\\
\hline
One-end trick			&	5.1 MeV		&	65	&	754	&		0.019		\\
One-end trick + TSM		&	4.9 MeV		&	409	&	809	&		0.019		\\
Time-dil.			&	13  MeV		&	31	&	745	&		0.126		\\
Time-dil. + TSM			&	7.5 MeV		&	281	&	710	&		0.040		\\
Time-dil. + HPE			&	8.0 MeV		&	34	&	750	&		0.048		\\
Time-dil. + HPE + TSM		&	6.2 MeV		&	322	&	750	&		0.029		\\
\hline
\multicolumn{5}{|c|}{$\sigma_{c}$ }					\\
\hline
One-end trick			&	95 MeV		&	65	&	144	&		1.30		\\
One-end trick + TSM		&	61 MeV		&	409	&	692	&		2.57		\\
\hline
\hline
\multicolumn{5}{|c|}{$g_A$ }						\\
\hline
One-end trick			&	0.19		&	65	&	2234	&		80.6		\\
One-end trick + TSM		&	0.081		&	409	&	1471	&		9.65		\\
\hline
\multicolumn{5}{|c|}{$g_A^s$ }						\\
\hline
One-end trick			&	0.076		&	65	&	754	&		4.36		\\
One-end trick + TSM		&	0.023		&	409	&	809	&		0.43		\\

Time-dil.			&	0.132		&	31	&	721	&		5.08		\\
Time-dil. + TSM			&	0.049		&	281	&	676	&		1.62		\\
Time-dil. + HPE			&	0.040		&	34	&	725	&		1.16		\\
Time-dil. + HPE + TSM		&	0.024		&	322	&	692	&		0.40		\\

\hline
\multicolumn{5}{|c|}{$g_A^c$ }						\\
\hline
One-end trick			&	0.076		&	65	&	144	&		0.83		\\
One-end trick + TSM		&	0.0215		&	409	&	692	&		0.32		\\
\hline 
\end{tabular}
\caption{Comparative computational cost for the $\sigma$-terms and axial charges using the different methods. The cost, in
  units of GPU-node seconds (2 GPUs per node), is given for the computation of the quark loop for one configuration, using
  $N_{\rm HP}=24$, $N_{\rm LP}=0$ and $N_{\rm HP}=8$ and $N_{\rm LP}=200$ or $N_{\rm LP}=300$ depending on the quark mass,
  as discussed above. For a fair comparison we used the same statistics, namely 18628 measurements, for time-dilution and
  the one-end trick. The sink was set at $t_s=16a$ for the one-end trick data, and the insertion to $t_{\rm ins}=8a$ for
  time-dilution. The column labeled as OH represents what we call the \emph{overhead}, in other words, the time of pre- and
  post-processing employed in generating the disconnected quark loops, whereas the cost includes the inversion time as well.
  It can be seen how the overhead time depends only on the number of sources calculated (not on the mass), and it becomes
  more and more important, as the matrix inversions become faster, that is, for larger quark masses. The last column
  defines a quantity that is independent of statistics, which gives the comparative cost for a fixed error of a given
  observable~\cite{Azcoiti:2009md}.  }\label{compAll}

\end{table}

\subsection{Summary on the performance of the various methods}

We summarize the outcome of the comparisons in Table~\ref{compAll}
where we give the computational cost and relative error for the
disconnected diagrams contributing to the $\sigma$-terms and the axial
charges for the light, strange and charm quarks.  A measure of the
comparative cost is given in seconds of computer time per
configuration on two Tesla M2070 GPUs.  Since all operator insertions
in the loop of a given quark flavor are computed simultaneously, the
cost for different observables is the same when using the same method.
To make the comparison meaningful, we restricted the number of
two-point functions used, so the statistics in all cases are 18628
measurements. The entry in the last column gives the comparative
advantage of each method~\cite{Azcoiti:2009md}.

From Table~\ref{compAll} it is clear that the one-end trick plus TSM
is the most suitable method for computing the disconnected
contributions to $\sigma_{\pi N}$ and $g_A$.  Since these observables
have very different convergence properties, we conclude that this
method will be preferable for the disconnected contributions due to
the light quark loops for other observables. For the strange quark
loops we have performed also a comparison with time-dilution. As can
be seen from the error$^2\times$cost, the one-end trick plus TSM is
also the preferred method over time-dilution plus any combination of
TSM and/or HPE.  For the charm quark loops the one-end trick performs
better as compared to including the TSM for $\sigma_c$.  However,
including the TSM clearly reduces the cost for a fixed error in the
case of $g_A^c$. Thus, since for a class of observables one needs to
use the TSM, using it also for the computation of $\sigma_c$ comes with no cost.

\section{Conclusions and outlook}
\label{sec:conclusions}

The computation of disconnected contributions for flavor singlet
quantities has become feasible, due to the development of new
techniques to reduce the gauge and stochastic noise, and due to the
increase in computational resources. In this work, we
explore a number of recent developments for the determination of
disconnected contributions to hadron matrix elements. The usage
of GPUs is particularly important, due to its efficiency in the
evaluation of disconnected diagrams using the TSM, since GPUs can
yield a large speedup when employing single- and half-precision for
the computation of the LP inversions and associated contractions.

Among all the algorithms analyzed, the one-end trick seems to perform better
in most cases, reducing the variance of the disconnected loops at the same
computational cost for many flavor-singlet quantities. It also delivers the
fermion loops for all the possible insertion times at no extra cost, so we can
use the summation method in the analysis, and the computation of one-derivative
insertions is straightforward, whereas for the case of time-dilution, several
separated inversions must be performed. 

The TSM can improve the efficiency of the one-end trick for quark
masses up to the strange quark mass.  For heavier masses, the
performance of the TSM degrades, and depending on the disconnected
quark loop to be computed it is no longer beneficial. In our case, we
observe a performance degradation for $\sigma_c$ but a clear
improvement for $g_A^c$  yielding results with smaller
errors. Thus for loops where the stochastic noise is expected to be
large the TSM still performs better even for heavy quark masses where
the CG converges fast.

In a follow-up paper we will apply the TSM to perform a high
statistics analysis of the disconnected diagrams involved in
observables probing nucleon structure.  These will include the
isoscalar electromagnetic and axial vector form factors, the
sigma-terms, the momentum fraction and helicity.

\section*{Acknowledgments}  
A. V. and M. C. are supported by funding received from the Cyprus
Research Promotion Foundation (RPF) under contracts EPYAN/0506/08 and TECHNOLOGY/$\Theta$E$\Pi$I$\Sigma$/0311(BE)/16 respectively. K. J. 
is partly supported by RPF under contract $\Pi$PO$\Sigma$E$\Lambda$KY$\Sigma$H/EM$\Pi$EIPO$\Sigma$/0311/16. This
research was in part supported by the Research Executive Agency of the
European Union under Grant Agreement number PITN-GA-2009-238353 (ITN
STRONGnet) and the infrastructure project INFRA-2011-1.1.20 number
283286 (HadronPhysics3), and the Cyprus RPF 
under contracts KY-$\Gamma$A/0310/02 and NEA
Y$\Pi$O$\Delta$OMH/$\Sigma$TPATH/0308/31 (infrastructure project
Cy-Tera, co-funded by the European Regional Development Fund and the
Republic of Cyprus through RPF). Computational resources were provided
by the Cy-Tera machine and Prometheus (partly funded by the EU FP7 project PRACE-2IP under grant agreement number: RI-283493)
of CaSToRC, Forge at NCSA Illinois (USA), Minotauro at BSC (Spain),
and by the Jugene Blue Gene/P machine of the J\"ulich Supercomputing
Center awarded under PRACE. 

\appendix

\setcounter{table}{0}
\setcounter{figure}{0}
\renewcommand\thetable{\Alph{section}.\arabic{table}}
\renewcommand\thefigure{\Alph{section}.\arabic{figure}}

\section{Details on the implementation of the GPU code}
\label{sec:app1}

\subsection{Twisted mass fermion operator}
In this section we provide some implementation aspects of the twisted mass code development in QUDA.  The Wilson twisted mass fermion operator formulation for the degenerate flavor doublet reads:\\
\begin{equation}
~~~\slashed{D}_{TM} = \slashed{D}_{W} + i \mu \gamma_5 \tau^{3},
\end{equation}
where $\slashed{D}_W$ stands for the Wilson term, $\tau^3$ denotes the diagonal $SU(2)$ Pauli matrix and $\mu$ is the (bare) twisted mass parameters. 
For internal computations QUDA adopts a non-relativistic basis  for  the spinor projections; this allows to
reduce memory traffic while computing hopping terms in time direction \cite{Clark:2009wm}. 

  For the QUDA twisted mass iterative solvers  one can employ two types of (even-odd) preconditioning: symmetric and asymmetric. For instance, one may deal with the following equivalent ('even-even') preconditioned systems:
\begin{equation}
(R_{ee} - \kappa^2 \slashed{D}_{eo} R^{-1}_{oo} \slashed{D}_{oe})\psi_{e} = b_e - \slashed{D}_{eo}R^{-1}_{oo} b_o~~~~~
\end{equation}
\begin{equation}
(I_{ee} - \kappa^2 R^{-1}_{ee} \slashed{D}_{eo} R^{-1}_{oo} \slashed{D}_{oe})\psi_{e} = R^{-1}_{ee}(b_e - \slashed{D}_{eo}R^{-1}_{oo} b_o)
\label{A.3}
\end{equation}
where $R$ represents a local twisting operator and the odd component of the solution is reconstructed by the expression:
\begin{equation}
\psi_{o} = R^{-1}_{oo}(b_o - \slashed{D}_{oe}R^{-1}_{ee} \psi_e). 
\end{equation}
Accordingly, we implemented a number of 'fused' CUDA kernels, such as
$~  R^{-1}_{oo} \slashed{D}_{oe}, (R_{ee} - \kappa^2 \slashed{D}_{eo})$ (and their 'daggered' analogues), required for the left-hand-side  of Eq. (A.2). As a result, all local operators are merged into dslash kernels and computed on the fly reducing expansive accesses to the GPU global buffer. All these kernels are generated by a python script in the same way as it is done for other fermion operators available in QUDA. 

Finally, to include the twisted mass dslash operator in the whole framework, we added two new classes, {\texttt{DiracTwistedMass}} and {\texttt{DiracTwistedMassPC}}, which encapsulate  all necessary attributes and methods for launching dslash kernels on the accelerators. The multi-GPU parallelization for the degenerate flavor doublet is almost identical to the corresponding Wilson implementation.  The only difference consists in the necessity to apply  the local twisting operator $R^{-1}$ (e.g., entering operators in the lhs of Eq.~(\ref{A.3}))  while gathering boundary-spinor sites: we provided with an extra packing routine to properly take into account this case.
More detailed information about optimization strategies exploited in the QUDA library can be found in Refs.~\cite{Clark:2009wm,Clark:2009qp,Babich:2011np}.

We adduce single-GPU performance summary for the asymetrically preconditioned dslash operator on the NVIDIA GTX Titan card. 
Here we included the plain Wilson case as a reference point. The lattice size for the single-GPU runs was $32\times64$ and
we examined two types of gauge field reconstructions, namely 8- and 12-parameter reconstructions.
QUDA allows for storing the gauge-field links in less than the 9 complex numbers needed to store
a full $SU(3)$ matrix. In one case, it allows omitting one row of the three, reducing the storage
requirements to 6 complex numbers, so-called 12-parameter reconstruction. With 8-parameter
reconstruction, the link is decomposed into a linear combination of the eight $SU(3)$ generators and
only the coefficients are stored (8 real numbers). In both cases the full $SU(3)$ is recomputed on the
fly during the Dirac operator application. This reduces both the memory requirements but more
importantly the bandwidth requirements of applying the Dirac matrix. In addition, to benefit from
full-clock speed for the double precision Arithmetic Logic Units on the gaming card we set

{\texttt{GPUDoublePrecisionBoostImmediate=1}}.

 We summarize our results  in Table~\ref{tab:single-gpu}. 
\begin{table}
\caption{Single GPU performance in GFlops.} \label{tab:single-gpu} 
\begin{tabular}{|c|c|c|c|}
  \hline
  Prec. & Recon. & Wilson & Degenerate TM\\
  \hline
  \hline
  double~~&~~12& 184~~~ & 190~~~~ \\
  ~~~~~~~~&~~8& ~179~~~~~ & ~~183~~~~~~\\
\hline
  single~~~&~~12& 401~~~ & 415~~~~\\
  ~~~~~~~~~&~~8& ~472~~~~ & ~~487~~~~~\\
\hline
  half~~&~~12& 732~~~ & 759~~~~\\
  ~~~~~~&~~8& ~829~~~~ & ~~858~~~~~\\
  \hline
\end{tabular}
\end{table}

\subsection{Contraction kernels}

A fundamental step in the calculation of quark loops  is the contraction of inverted sources.
To this end we developed efficient GPU code yielding $\sim300$ GFlops in a single Tesla m2070 GPU in
double precision, and showing almost perfect scaling with increasing number of GPUs.

Traces were taken in color space, leaving the Dirac and volume indices open. The volume indices are
used later for the FFT, so we obtain solutions for different momenta, whereas the open Dirac indices
are there in order to deal with the different insertions. We calculated the outer product in Dirac
space of both sources to be contracted, and consequently a $4\times4$ matrix was obtained, with enough
information to reconstruct any arbitrary $\gamma$ insertion just by transposition and multiplication.
Therefore, our contraction code automatically outputs all the possible insertions for ultra-local
operators. A covariant-derivative kernel was also developed to allow the calculation of one-derivative
insertions.

These GPU kernels were developed in two different versions: whole spinor contraction (for the one-end trick)
and single time-slice contraction (for time-dilution). The single time-slice contraction kernel does not
support at this moment the inclusion of the covariant-derivative, for this would imply to deal with several
time-slices at the same time.

\subsection{Interfaces and workflow}

Since QUDA already implements most of the code we need for computing disconnected diagrams, the largest
contribution to the library on this package is the writing of interfaces. Those were designed to calculate
any ultra-local and one-derivative insertion with several variance reduction methods, namely the TSM, the
one-end trick (only for twisted mass fermions), time-dilution and the Hopping Parameter Expansion,
and all the possible combinations of these.

The interface generates random stochastic sources using RANLUX from the GSL library on the CPUs; then the
source is sent to the GPUs for inversion and contraction, and contractions are stored back in the CPUs. This
process is repeated several times for the binary storage system, explained below. After we accumulated enough
sources, the data is sent back from CPUs to GPUs for FFT using the NVIDIA library cuFFT; our output from the last
section fits exactly in the input required in the functions of the cuFFT library, so no further transformations are
required. At this point, all possible momenta are generated, but we usually insert here a cut-off in $p^2$ to reduce
storage. Finally, the results are written to disk in a parallel fashion to reduce I/O time.

\subsection{Binary storage system}

When the TSM is introduced, one has to face the problem of storage and take decisions regarding which
information can be discarded, due to the large number of stochastic sources generated. In our case
it was decided that only contractions would be stored, but even in this case a storage problem might
appear.

For instance, if we decide to use volume sources in our calculation (i.e. we are using the one-end trick),
our code will compute all ultra-local and one-derivative insertions, with and without a flavor $\tau_3$ matrix,
that is, 160 insertions in total. We impose a momentum cut-off that we can set in our example to $p^2 <= 9$. In this case, each configuration with
volume $32^3\times 64$ takes around $\sim50$Gb storage, a huge number taking into account that GPU applications
are usually not granted as much disk storage as CPU's. In the GPU clusters we run our code, we were granted between
$5$ and $20$Tb of disk space, which we would fill with 100-400 configurations, a number that might hardly be enough
for a single ensemble, let alone when dealing with several ensembles.

An obvious way to reduce storage needs is to transform the data from text to binary format, which
will grant us a $\sim70\%$ reduction in disk usage, although it will be still a huge amount of data.
A further reduction of storage requirements is only achieved through clever techniques; in our
case we developed a binary-storage technique, inspired in the way the bits make up a byte. This
technique reduced storage requirements up to $\sim97\%$ without losing any relevant information.

\begin{figure}[h!]
\hspace{1.5cm}File.1, File.2, File.4, File.8
\newline
\setlength{\unitlength}{1cm}
\begin{picture}(8,1)
\put(2.7,1){\line(1,-1){1}}
\put(3.8,1){\line(2,-5){0.4}}
\put(4.9,1){\line(-2,-5){0.4}}
\put(6.0,1){\line(-1,-1){1}}
\end{picture}
\newline
\begin{tabular}{rl}
Weight: & \framebox{1}\framebox{2}\framebox{4}\framebox{8} \\
Mask: & \framebox{1}\framebox{0}\framebox{1}\framebox{1} \\
\hline
Total: & \framebox{1}\framebox{0}\framebox{4}\framebox{8} = 13 \\
\end{tabular}
\caption{Example of construction of the inverse estimator using 13 sources in our storage
method.\label{St}}
\end{figure}
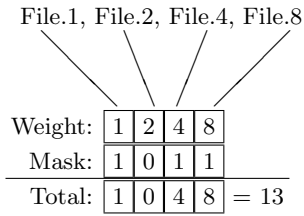

In order to understand the way the technique works, we can have a look at Fig.~\ref{St}. The byte
is composed of bits, and each bit has a different wait according to its position (1, 2, 4\ldots).
The idea is to mimic this structure for the stochastic sources. Since in the end we are going
to average the sources, we can add several and store the addition in a single file; so in the file File.1 we store the contractions generated with one stochastic source, in File.2 we store the sum
of the contractions coming from the second and the third stochastic source, and so on. Reconstruction
is straightforward, taking into account the base-2 structure.

With this storage method one can recover the data for any number of sources, therefore we are
keeping the same information in much less space. Actually some information is lost, for after
storing the data there is only one way to recover a fixed number of sources, whereas before 
there could be many, but this extra information is not useful for us and can be discarded. In
contrast, we gain a huge reduction in storage requirements, from $O\left(N\right)$ to
$O\left(\log_2 N\right)$.

\bibliography{ref}

\end{document}